\documentclass[amsmath,eqsecnum,preprint,pre,aps,showpacs]{revtex4}
\usepackage{amsmath}
\usepackage{bm}
\usepackage{graphicx}
\usepackage{xcolor}

\begin{document}

\title{Non-equilibrium  Onsager-Machlup theory}
\author{Ricardo Peredo-Ortiz$^1$, Luis F. Elizondo-Aguilera$^2$, Pedro Ram\'irez-Gonz\'alez$^1$, Edilio L\'azaro-L\'azaro$^1$, Patricia Mendoza-M\'endez$^3$, and Magdaleno Medina-Noyola$^1$}

\address{1 Instituto de F\'{\i}sica {\sl ``Manuel Sandoval Vallarta"}, Universidad Aut\'{o}noma de San Luis Potos\'{\i}, \'{A}lvaro Obreg\'{o}n 64, 78000 San Luis Potos\'{\i}, SLP, M\'{e}xico}
\address{2 Instituto de F\'isica, Benem\'erita Universidad Aut\'onoma de Puebla, Apartado Postal J-48, CP 72570 Puebla, M\'exico}
\address{3 Facultad de Ciencias F\'isico-Matem\'aticas, Benem\'erita Universidad Aut\'onoma de Puebla, Apartado Postal 1152, CP 72570 Puebla, M\'exico}
\date{\today}

\begin{abstract}

This paper proposes a simple mathematical model of non-stationary and non-linear stochastic dynamics, which approximates a  (globally) non-stationary and non-linear stochastic process by its locally (or \emph{``piecewise''}) stationary version. Profiting from the elegance and simplicity of both, the exact mathematical model referred to as the Ornstein-Uhlenbeck stochastic process (which is \emph{globally} stationary, Markov and Gaussian) and of the Lyapunov criterion associated with the stability of stationarity, we show that the proposed non-linear non-stationary model provides a natural extension of the Onsager-Machlup theory of equilibrium thermal fluctuations, to the realm of non-stationary, non-linear, and non-equilibrium processes. As an illustrative application, we then apply the extended non-equilibrium Onsager-Machlup theory, to the description of thermal fluctuations and irreversible relaxation processes in liquids, leading to the main exact equations employed to construct the \emph{non-equilibrium self-consistent generalized Langevin equation} (NE-SCGLE) theory of irreversible processes in liquids. This generic theory has demonstrated that the most intriguing and long-unsolved questions of the glass and the gel transitions are understood as a natural consequence of the second law of thermodynamics, enunciated in terms of the proposed piecewise stationary stochastic mathematical model.
\end{abstract}


\maketitle

\section{Introduction.}\label{section1}


Some fundamental physical laws often seem simpler and most elegant when expressed in the right mathematical language and conceptual perspective, as illustrated by Newton's second law of mechanics when expressed in the language of differential calculus or with the use of generalized coordinates and canonical transformations \cite{goldstein}. Similarly, specific symmetries of physical systems and processes allow the identification of exact conservation laws, as illustrated by the application of Noether's theorems \cite{noether,hermannschmidt}. Probability and statistics, on the other hand, express the laws of macroscopic physics in its most precise mathematical terms, summarized by the fundamental principles of statistical mechanics \cite{huang,mcquarrie}.  

Many physical and natural phenomena that involve the time-evolution of random variables are most naturally described in the language of stochastic processes  \cite{keizer,coffey2004}. In fact, the initial development of this mathematical infrastructure was intimately related with the need to understand specific physical phenomena \cite{wax}. Thus, in his description of Brownian motion, Langevin modeled the random movement of a given colloidal particle, by applying Newton's second law to the particle, subjected to an approximately linear hydrodynamic resistance force, plus a fluctuating force caused by thermal fluctuations \cite{langevin}. Thus, the particle's random velocity was modeled as the solution of a linear differential equation with an additive random term, assumed to be \emph{``indifferently positive and negative and with a magnitude such that it maintains the agitation of the particle, which the viscous resistance would stop without it"}  \cite{langevin}, a physically intuitive manner employed by Langevin to express his assumption that such random term is a \emph{stationary} $\delta$-correlated noise (i.e., a ``white'' noise). From this definition, all the mathematical properties (i.e., all the moments and correlations of all the pertinent distribution functions) can be determined exactly \cite{keizer}, thus showing that the resulting process is a Gaussian, stationary, and Markov process. Such a mathematical model is referred to as an Ornstein-Uhlenbeck (OU) stochastic process \cite{ornsteinuhlenbeck,wanguhlenbeck}. In this manner, without explicitly declaring it, Langevin introduced for the first time this simple and exactly solvable mathematical model, which is also a cornerstone of the mathematical field of stochastic processes \cite{wax}.

The same mathematical model, however, soon found a still more general application in physics, in Onsager's theory of transport and thermal fluctuations. For this we mean Onsager's laws of linear irreversible thermodynamics (LIT) \cite{onsager1, onsager2} and the corresponding stochastic theory of thermal fluctuations by  Onsager and Machlup \cite{onsagermachlup1,onsagermachlup2,machlup0}. In essence, the main postulate of the Onsager-Machlup theory is the physical assumption that the thermal fluctuations of the extensive thermodynamic variables of any equilibrium system are described by the same mathematical model that describes the velocity of Langevin's Brownian particle, i.e., by the Ornstein-Uhlenbeck stochastic process. Among the many applications of the Onsager-Machlup theory, it has been established \cite{fox1,fox2}, for example, that the Landau-Lifshitz stochastic version of the hydrodynamic equations \cite{landaulifshitzfm} and the linearized Boltzmann equation with added fluctuations \cite{fox2,bixonzwanzig}, are two concrete examples of fluctuation theories that fit in such general scheme \cite{keizer}.


Although the Ornstein-Uhlenbeck stochastic process is a beautiful, simple, and exact mathematical model, its Markov and Gaussian nature severely limits the applicability of the original Onsager-Machlup theory of thermal fluctuations. For example, the assumption that thermal fluctuations are always Gaussian is in fact not accurate, as evidenced by the fact that at least the exact equilibrium probability distribution function $W^{eq}[{\textbf a }] = e^{\left( S[{\textbf a }]-S[{\textbf a }^{eq}] \right)/k_B}$ (where $S$ is the entropy) is in general non-Gaussian \cite{callen}. Eliminating this limitation is easily achieved by excluding the unnecessary assumption of Gaussianity from the original Onsager-Machlup theory; this will still maintain virtually all of its most relevant and fundamental properties.

Besides Gaussianity, however, there are also reasons to remove the other explicit mathematical limitation of the Onsager-Machlup theory, namely, the assumption that thermal fluctuations are always Markovian. For example, memory effects, inherent in many physical phenomena (such as the diffusive properties of concentrated colloidal dispersions \cite{naegelereview}) provide evidences of the contrary. The possibility of including memory, however, may be readily implemented if the supporting mathematical model (the OU process) is replaced by a more flexible model stochastic process, whose only essential attribute is its stationarity, but which is no longer necessarily Markov nor Gaussian. Such a mathematical model was discussed in Refs. \cite{delrio,faraday}, which also demonstrated a surprising and relevant fact, namely, that the corresponding linear stochastic equation \emph{with memory} and additive (no longer white) noise, turns out to have the same mathematical structure as the so-called \emph{generalized Langevin equation} (GLE) \cite{boonyip}. The GLE was a widely used concept in the early description of thermal fluctuations in simple liquids \cite{copleylovesey,boonyip}, although in strong association with the Mori-Zwanzig projection operator formalism \cite{kubo,zwanzig,mori,berne}. Ref. \cite{delrio}, however, makes it clear that the mathematical structure of the GLE is a consequence of stationarity, and not of the Hamiltonian basis of its Mori-Zwanzig's derivation. 

In summary, the Onsager-Machlup theory has two severe self-imposed limitations: assuming that thermal fluctuations are Gaussian and Markovian. However, removing these limitations was more of a technical detail, as they only refer to the underlying mathematical framework. In contrast, the theory's major limitation refers to the \emph{physical} restriction to describe thermal fluctuations around \emph{thermodynamic equilibrium} states. As recently explained in Ref.  \cite{OUgranular}, there seems to be no fundamental impediment to using the Ornstein-Uhlenbeck stochastic process also to describe fluctuations in \emph{non-equilibrium stationary }systems. This clearly suggests a route to extend the Onsager-Machlup theory, including its non-Gaussian and non-Markov versions, to systems that are not in thermodynamic equilibrium. 

In preparation for such exploration, let us mention here that the hydrodynamic description of the spatio-temporal distribution and transport of mass, energy, momentum, and other extensive thermodynamic variables of a fluid \cite{landaulifshitzfm}, constitutes the most paradigmatic application of the original Onsager-Machlup theory \cite{degrootmazur,keizer}. Thermodynamic transport, however, is expressed in the language of non-linear deterministic differential equations, whose stationary solutions and their stability are the subject of careful mathematical analysis \cite{keizer}, involving the existence of a Lyapunov function \cite{lyapunov,parks} as a signature of stability. As explained by Prigogine  \cite{prigogine0}, in conventional hydrodynamics, such Lyapunov function is essentially (the negative of) the thermodynamic entropy.  

 In fact, the corresponding hydrodynamic equations may also have stationary solutions representing non-equilibrium stationary states of \emph{open systems} \cite{chandrasekharhhs}. However, many of the most spectacular manifestations of these non-equilibrium stationary states, such as the Rayleigh-B\'enard stationary convection cells (and any of what Prigogine \cite{prigogine0,nicolisprigogine} refers to as dissipative structures), fade away when the input and output of matter and energy are halted, and the system, now closed,  is allowed to eventually reach its final stationary state. Since the hydrodynamic equations conform to the original Onsager-Machlup theory, these final stationary states are always thermodynamic equilibrium states. The universal catalog of thermodynamic equilibrium states of matter is formed by all the states that maximizes the entropy, i.e., that are the solution of the equation $dS[\mathbf{A}]=0$, where $S$ is the total entropy (including reservoirs) and the components of the vector $\mathbf{A}$ are the extensive thermodynamic variables \cite{callen}.

Unfortunately, conventional hydrodynamics, and in particular Prigogine's theory of dissipative structures, leave out many relevant processes in physical systems, which even in full isolation, attain final stationary states that do not maximize the entropy, and hence, are not contained in the universal catalog of  thermodynamic equilibrium states. We refer to otherwise extremely common kinetically-arrested materials, such as  glasses, gels, solid foams, and many other non-equilibrium amorphous solid materials, whose formation exhibits an extremely slow relaxation kinetics (``aging''), and whose properties depend on the protocol of fabrication \cite{angellreview1,ngaireview1,anderson,sciortino}. 

Since the formation of these materials is the result of  intrinsically non-stationary and non-linear processes,  a major theoretical challenge of contemporary statistical physics consists of endowing hydrodynamics with the kinetic pathways to the alternative universe of kinetically-arrested stationary  states of matter. One possible route may be to explore further extensions, now to non-stationary, non-equilibrium, and non-linear conditions, of the Onsager-Machlup theory. Identifying such an extension is, in its own right, another long-standing challenge of statistical physics \cite{keizer,casasvazquez0,casasvazquez1}. 

In this context, let us mention that both major statistical physics challenges (i.e., extending Onsager's theory to non-equilibrium conditions, and deriving from such an extension a first-principles statistical mechanical
 microscopic description of the amorphous solidification of liquids), have been successfully addressed  by the Mexican statistical physics community within the last two  decades. Since the mathematical condition of stationarity \cite{delrio}, and not the physical condition of thermodynamic equilibrium,  is the essence of  the Onsager-Machlup  formalism (including the GLE), it is natural to consider  modeling a \emph{non-stationary} stochastic process as a piecewise sequence of stationary processes. This argument led to the proposal of a non-equilibrium extension of the Onsager-Machlup theory, including a non-stationary version of the GLE \cite{nescgle00}. Such general formalism provided  a route to the description of non-equilibrium and non-stationary processes \cite{nescgle0}. 
 
It was precisely along this route, that the application of such an abstract time-dependent GLE canonical formalism, gave rise in 2010 to  the groundbreaking development of a generic first-principles theory of the  non-equilibrium structure and dynamics of liquids, referred to as the \emph{non-equilibrium self-consistent generalized Langevin equation} (NE-SCGLE) theory, first reported in the PhD thesis of Pedro Ram\'irez-Gonz\'alez \cite{tesispedro} and in the subsequent publications \cite{nescgle1,nescgle2,nescgle3}. During the last two decades, the solution of the non-linear time-evolution equations that constitute the core of this theory, unveiled an amazing predicted scenario, involving two competing  kinetic pathways: the processes of thermodynamic equilibration, and the ultra-slow dynamic arrest processes leading to the formation of non-equilibrium amorphous solids. The stationary  solutions of these time-evolution equations are thus naturally grouped in two mutually-exclusive sets: the well-understood universal catalog  formed by all the states that maximize the entropy, and the newly-discovered, complementary universal catalog, of non-equilibrium amorphous states of matter. 


The NE-SCGLE theory's predictions thus offer a fresh and accurate explanation of the behavior of glasses, gels, and other amorphous solids that exhibit dynamic arrest, including the corresponding transitions. Unfortunately, these discoveries have received only modest attention from the international scientific community, partly due to a natural and healthy skepticism toward unexpected solutions of long-standing problems. For example, it may be challenging to accept that the fundamental understanding of a complex problem like the glass transition, can be understood from the apparently simple kinetic perspective provided by the non-equilibrium extension of the Onsager-Machlup theory. It is thus necessary to develop new convincing  applications of the NE-SCGLE theory, while simultaneously revisiting and strengthening its conceptual foundations. 

Along the latter intention, the present work is aimed at elaborating a simpler and logically more economical statement of the fundamental physical assumptions of the non-equilibrium  Onsager-Machlup  theory. For this, here we update and enrich its underlying mathematical framework, by merging the original definition of stationary non-Markov stochastic processes \cite{delrio} with elementary notions of Lyapunov stability of stationary states, thus clarifying the original efforts to define the simple mathematical model of non-stationary stochastic process, referred to as piecewise stationary (PS),  which frames our non-equilibrium extension of the Onsager-Machlup theory to full non-linear and non-equilibrium conditions. In addition, we illustrate the use o this general formalism with a brief description of its application to the derivation of the  NE-SCGLE theory.

Our presentation starts in Section \ref{section2} with a brief review of the most fundamental physical principles  (the Langevin theory of Brownian motion and the Onsager-Machlup theory of thermal fluctuations) that allowed the development in the year 2000, of the precursor of the  NE-SCGLE theory, i.e., the first-principles theory of the dynamic properties of liquids \emph{in thermodynamic equilibrium}, referred to simply as the  \emph{Self-Consistent Generalized Langevin Equation} (SCGLE) theory (without the NE- part of NE-SCGLE). This equilibrium theory, which was presented in the PhD thesis of Laura Yeomans-Reyna \cite{tesislaura}  and in later publications \cite{scgle1,scgle2,scgle3,scgle4},  turned out to exhibit a spectacular predictive power. Furthermore, it happened to be equivalent in most respects to the well-established mode coupling theory  (MCT) of the glass transition \cite{goetze1,goetze2,goetze3,goetze4,mctjenssen}.


Section \ref{section3} introduces the mathematical model of a globally non-stationary (but locally stationary) stochastic process, referred to as \emph{piecewise stationary (PS)}  process, defined by two basic assumptions. The first concerns the mean value of the stochastic process, which is assumed to be determined by a deterministic non-linear differential equation, and the second assumes a particular time-evolution stochastic equation for the fluctuations around this mean value. From this definition a number of general properties of the PS stochastic process can be derived, the most important of which are also listed in Section \ref{section3}. 

This mathematical framework is then employed in Section \ref{section4} to enunciate the main postulates of the non-equilibrium, non-stationary, and non-linear extension of the Onsager-Machlup theory \cite{nescgle0}. The application of this general and abstract theoretical framework to liquids, then leads to the generic first-principles non-equilibrium theory of the structure and dynamics of liquids referred to as the NE-SCGLE theory, whose derivation is briefly reviewed in Section \ref{section5}. This work ends with a short summary in Section \ref{section6}.

\section{Background and fundamental physical principles.}\label{section2}

As said in the introduction, the initial development of the concept of stochastic processes as a mathematical infrastructure was intimately related with the need to understand specific physical phenomena \cite{wax}. Thus, Langevin modeled the Brownian movement of a given colloidal particle, by applying Newton's second law for the particle, subjected to an approximately  linear hydrodynamic resistance force $-\zeta\textbf{V}(t)$, where $\mathbf{V}(t)$ is the particle's random velocity (and $\zeta$ its friction coefficient), plus a fluctuating force caused by thermal fluctuations \cite{langevin}. Thus, $\mathbf{V}(t)$ was modeled as the solution of a linear differential equation with an additive random term, namely,
\begin{equation}
M\frac{d\textbf{V}(t)}{dt}= -\zeta\textbf{V}(t)+\textbf{f}(t),
\label{ordlangevin}
\end{equation}
in which $M$ is the mass of the particle and the added term $\textbf{f}(t)$ represents the random  thermal fluctuating force exerted by the supporting solvent, assumed to be a Gaussian \emph{stationary} $\delta$-correlated noise. 

From this definition,  all the mathematical properties (i.e., all the moments and correlations of all the pertinent distribution functions) can be determined exactly \cite{keizer}, thus showing that the resulting process $\mathbf{V}(t)$ is a Gaussian, stationary, and Markov process, eventually referred to as an Ornstein-Uhlenbeck (OU) stochastic process \cite{ornsteinuhlenbeck,wanguhlenbeck}.

\subsection{The Ornstein-Uhlenbeck (OU) stochastic process.}\label{subsection2.1}

Mathematically, an Ornstein-Uhlenbeck stochastic process is described by an $N$-component stochastic vector $\textbf{a}(t)$ with components  $a_i(t)$ (with $i=1, 2, ...., N$) and mean value ${\overline {\textbf a }}^{ss}$,  defined by the solutions of the following linear stochastic differential equation for the fluctuations $\delta {\textbf a }(t)\equiv {\textbf a }(t)-{\overline {\textbf a }}^{ss}$,
\begin{equation}
\frac{d \delta {\textbf a }(t)}{dt}=   \mathcal{H} \cdot \delta
{\textbf a }(t) +{\textbf f}(t), \label{ddeltaadt}
\end{equation}
with $\mathcal{H}$ being a $N\times N$ relaxation matrix and with the $N$-component vector ${\textbf f}(t)$ being a Gaussian ``white noise", i.e., a stationary Gaussian stochastic process with zero mean ($\overline{{\textbf f}(t)}={\textbf 0}$) and $\delta$-correlated, ($\overline{\textbf{f}(t)\textbf{f}^{T}(t')}= {\bm{\gamma}} 2\delta (t-t')$).

From this definition,  all the statistical properties of the stochastic vector $\textbf{a}(t)$ can be derived \cite{keizer}. The exact solution consists of analytic expressions  for all these properties in terms of the matrix $\textbf{H}$ and the stationary covariance ${\bm{\sigma}}^{ss}\equiv <\delta {\textbf a }(t)\delta {\textbf a }^\dagger(t)>^{ss}$. This includes the fluctuation-dissipation relation  $\mathcal{H}\cdot {\bm{\sigma}}^{ss} + {\bm{\sigma}}^{ss} \cdot  \mathcal{H}^{T } +2{\bm{\gamma}} =0$ for the matrix ${\bm{\gamma}}$, and the closed expressions for the conditioned mean value, $\langle \mathbf{a}(t) \rangle^{0}= {\overline {\textbf a }}^{ss}+ e^{-\mathcal{H}t}\cdot [\mathbf{a}^0- {\overline {\textbf a }}^{ss}]$, and for the conditioned covariance ${\bm{\sigma}}(t)\equiv \langle\delta \mathbf{a}(t)\delta \mathbf{a}^\dagger (t') \rangle^0={\bm{\sigma}}^{ss} \cdot [1-e^{-\mathcal{H}t}]$. The stationary two-times correlation function, $\textbf{C}(\tau)\equiv \langle \delta \mathbf{a}(t) \delta\mathbf{a}(t+\tau) \rangle^{ss}$  is given by $\textbf{C}(\tau)= e^{-\mathcal{H} \mid \tau \mid}\cdot\mathbf{\sigma}^{ss}$. Finally, one can derive a handy exact expression for the second moment $\textbf{W}(t)\equiv <\Delta \textbf{x}(t)\Delta \textbf{x}^\dagger(t)>$ of the ``displacement" $\Delta \textbf{x}(t)\equiv\int_{0}^{t}\textbf{a}(\tau)d\tau$. For the mono-component case ($\nu=1$), such an expression reads $W(t)=2H^{-1}\sigma^{ss}[t+H^{-1}(e^{-Ht}-1)]$, which  at short times is quadratic and at long times is linear in the time $t$. Notice that, in  spite of the physical flavor of the name ``fluctuation-dissipation relation'' and  the concept of ``displacement" $\Delta \textbf{x}(t)$ (and ``mean square displacement'' for  $\textbf{W}(t)$),  these are, in reality,  purely mathematical concepts.

Today we can use this mathematical infrastructure to describe Langevin's theory of Brownian motion more efficiently. This can be done in terms of two fundamental postulates. The first is that the random character of the particle's velocity $\textbf{V}(t)$ is described by an Ornstein-Uhlenbeck stochastic process with vanishing stationary mean value, $\overline{\textbf{V}}^{ss}=  \textbf{0}$, and  specified by the diagonal relaxation matrix $\mathcal{H}=-(\zeta/M)\textbf{I}$ and by the stationary covariance  ${\bm{\sigma}}^{ss}\equiv \overline{\textbf{V}(t)\textbf{V}(t)}$.  From this postulate, all the properties of the Ornstein-Uhlenbeck stochastic  $\textbf{a}(t)$ apply for the velocity $\textbf{V}(t)$ of the  Brownian particle. 

Notice that all these properties  are a consequence of the condition of stationarity, and NOT of the condition of thermodynamic equilibrium. Thus, their applicability is not restricted by the equilibrium condition, but only by the condition of stationarity, as discussed in Ref.  \cite{OUgranular}. To restrict the theory to equilibrium conditions, an additional postulate is introduced, which determines the stationary covariance  ${\bm{\sigma}}^{ss}\equiv \overline{\textbf{V}(t)\textbf{V}(t)}$ by the equipartition theorem of  statistical thermodynamics,   $\overline{\textbf{V}(t)\textbf{V}(t)}= (k_BT/M)\textbf{I}$ (with $\textbf{I}$ being the $3\times 3$ unit matrix). In this manner, one can easily check that most textbook presentations of Langevin's theory of Brownian motion \cite{mcquarrie,chandrasekhar} then become essentially a review of the physical meaning  of each mathematical property of the underlying Ornstein-Uhlenbeck mathematical model.

\subsection{From Langevin to Onsager: same mathematical model.}\label{subsection2.2}

The Ornstein-Uhlenbeck mathematical model soon found a still more general application in physics, in Onsager's theory of transport and thermal fluctuations. For this we mean Onsager's laws of linear irreversible thermodynamics (LIT) \cite{onsager1, onsager2} and the corresponding stochastic theory of thermal fluctuations by  Onsager and Machlup \cite{onsagermachlup1,onsagermachlup2,machlup0}. In essence, the main postulate of the Onsager-Machlup theory is the physical assumption that the thermal fluctuations of the extensive thermodynamic variables $a_i(t)$ (with $i=1, 2, ...., N$)  of any equilibrium system are described by the same mathematical model that describes the velocity of Langevin's Brownian particle, i.e., by the Ornstein-Uhlenbeck stochastic process \cite{keizer}. 

As mentioned before, the Gaussian and Markov nature of the Ornstein-Uhlenbeck stochastic process severely limits the applicability of the original Onsager-Machlup theory of thermal fluctuations. For example, the exact equilibrium probability distribution function $W^{eq}[{\textbf a }] = e^{\left( S[{\textbf a }]-S[{\textbf a }^{eq}] \right)/k_B}$ (where $S$ is the entropy) is in general non-Gaussian \cite{callen}. However, eliminating  Gaussianity from the original Onsager-Machlup theory is a minor technical detail since the non-Gaussian OU process will still maintain its most relevant and fundamental properties. Similarly,  the memory effects inherent, for example, in the diffusive properties of concentrated colloidal dispersions \cite{naegelereview}), evidence the limitation imposed by the Markov condition. This, however, is also more of a technical detail, as discussed in what follows.

\subsection{The fundamental role of stationarity: Mori-Zwanzig vs. Onsager with memory.}\label{subsubsection2.3}

The possibility of including memory in the Onsager-Machlup theory may be readily implemented if the supporting mathematical model (the OU process) is replaced by a more flexible model stochastic process, whose only essential attribute is its  stationarity, but which is no longer necessarily Markov nor Gaussian. Such a mathematical model is now defined by a linear stochastic equation \emph{with memory} and additive noise,
\begin{equation}
\frac{d \delta {\textbf a }(t)}{dt}=   \int_0^t {\textbf H}(t-t') \cdot \delta {\textbf a }(t') dt' +{\textbf f}(t),
\label{ddeltaadtnonmarkov}
\end{equation}
where ${\textbf H}(t-t') $ is a  $N\times N$ matrix of memory functions, and  with the additive noise ${\textbf f}(t)$ no longer  necessarily Gaussian nor $\delta$-correlated, \emph{but  necessarily stationary}, with zero mean, $\overline{{\textbf f}(t)}={\textbf 0}$, and time correlation function given by $\overline{{\bf f}(t){\bf f}^{\dagger}(t')}=  {\bm{\Gamma}}(t-t')$. 

This mathematical model was discussed in Refs. \cite{delrio,faraday}, which demonstrated a surprising and fundamental fact, referred to as the \emph{stationarity theorem}. It states that the condition of stationarity (i.e., the invariance with respect to the origin of the time $t$), is a necessary and sufficient \emph{mathematical} condition for Eq. (\ref{ddeltaadtnonmarkov}) above, to be written in the very stringent and rigid mathematical format,  referred to as the \emph{generalized Langevin equation} (GLE), namely, 
\begin{equation}
\frac{d\delta  {\textbf a }(t)}{dt}= - {\bm{\omega}} \cdot {\bm{\sigma}}^{ss-1}
\cdot \delta  {\textbf a }(t) - \int_0^{t}dt'  {\bm{\Gamma}}(t-t')\cdot
{\bm{\sigma}}^{ss-1}\cdot \delta {\textbf a }(t')+{\textbf f }(t),
\label{gle0}
\end{equation}
with ${\bm{\omega}}$ being an antisymmetric matrix and with the matrix $ {\bm{\Gamma}}(t)$ having the symmetry imposed by its definition $ {\bm{\Gamma}}(t) =  {\bm{\Gamma}}^{\dagger}(-t)= {\overline {{\textbf f }(t){\textbf f }^{\ \dagger}(0)}}$.  

Many readers, however, may immediately recognize the GLE in Eq. (\ref{gle0}) as the result of a statistical mechanical  derivation that starts at the level of the (Newtonian or Hamiltonian) description of the  microscopic $N$-particle dynamics. Indeed, the GLE was a widely used concept in the early description of thermal fluctuations in simple liquids \cite{copleylovesey,boonyip}, where it became strongly associated with the Mori-Zwanzig projection operator formalism \cite{kubo,zwanzig,mori} employed in its derivation. However, Ref. \cite{delrio} makes it clear that the mathematical structure of the GLE is a consequence of stationarity, and not of the Hamiltonian basis of its Mori-Zwanzig's derivation. 
Of course, regarding the applications of the GLE above, to the description of fluctuations in equilibrium systems, the difference in the theoretical derivation of Eq. (\ref{gle0}) (Mori-Zwanzig vs. Onsager with memory) is completely irrelevant. In fact, we now describe one such application. 

\subsection{The equilibrium \emph{Self-Consistent Generalized Langevin Equation} (SCGLE) theory.}\label{subsection2.4}

\begin{figure*}[ht]
\centering
\includegraphics[width=8cm]{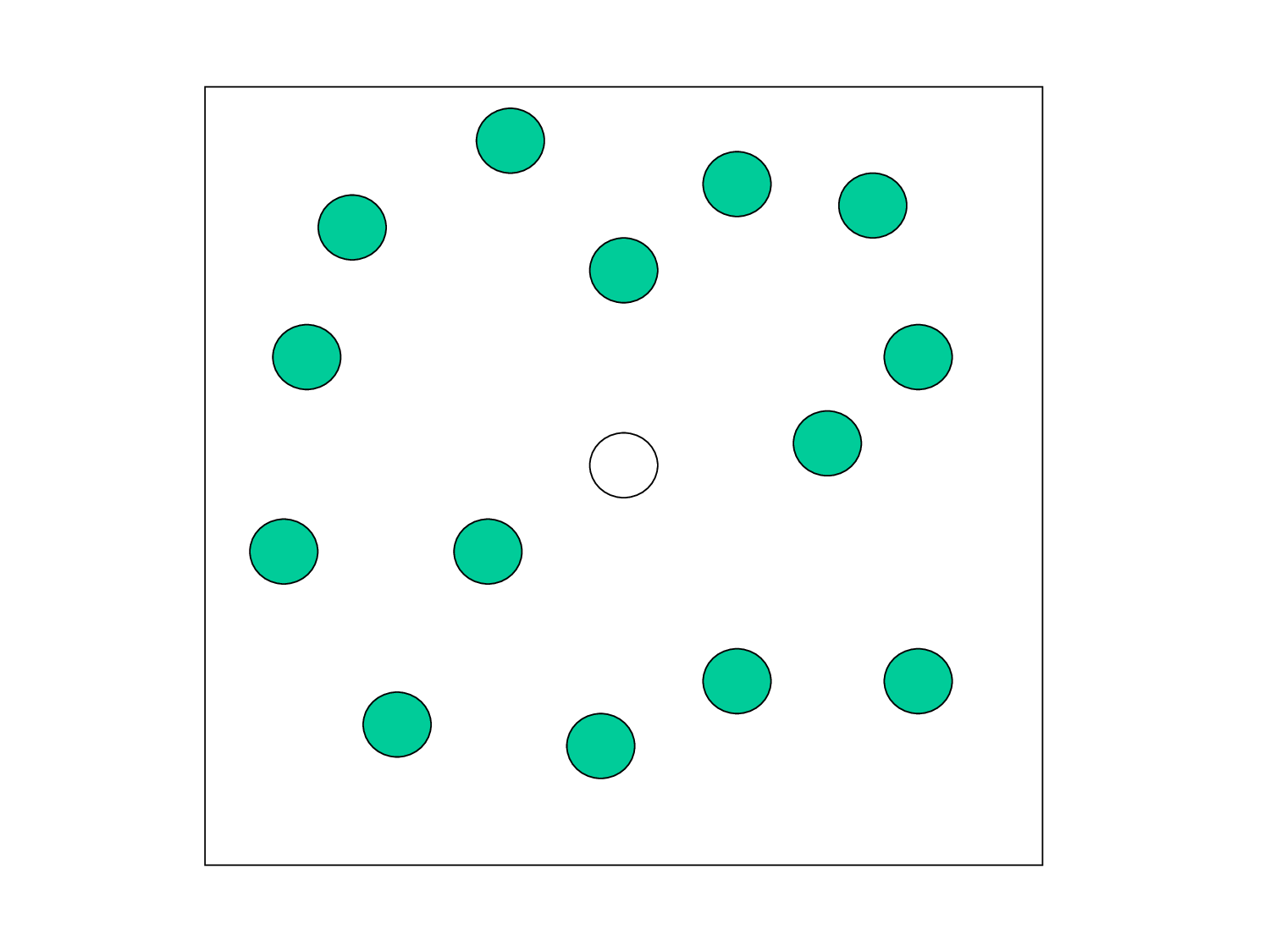}
\caption{A labelled tracer Brownian particle (empty circle) interacting with the surrounding particles (filled circles) of a colloidal suspension.}
\label{tracer}
\end{figure*}

The stationarity theorem \cite{delrio} was, thus, a fundamental cornerstone to ``liberate'' the Onsager-Machlup theory of thermal fluctuations from its  limitations (being Gaussian and  Markov) imposed by the Ornstein-Uhlenbeck mathematical model. In parallel with its publication in Ref.  \cite{delrio}, this extended version of the Onsager-Machlup theory had its first application in a very concrete problem, namely, the description of the effects on the motion of a tracer Brownian particle, caused by its direct interactions with the other Brownian particles in a concentrated colloidal suspension Ref. \cite{faraday}. The stationarity theorem and another related mathematical result (referred to as \emph{the contraction  theorem}  \cite{delrio}), were employed in  Ref. \cite{faraday} to write the Langevin equation for the velocity $\mathbf{V}(t)$ of such tracer particle (empty circle in Fig. \ref{tracer}),  coupled through the direct inter-particle  forces, with the fluctuations $\delta n(\mathbf{r},t)$ of the  local concentration of the surrounding colloidal particles (solid circles).

The mathematical constraints and selection rules of the matrices   ${\bm{\sigma}}^{ss}$,  ${\bm{\omega}}$, and  ${\bm{\Gamma}}(t)$ of the GLE in Eq. (\ref{gle0}), applied to the stochastic vector  $\delta  {\textbf a }(t)\equiv [\mathbf{V}(t),\delta n(\mathbf{r},t)]$, allow writing the stochastic diffusion equation for  the fluctuations $\delta n(\mathbf{r},t)$. These two stochastic equations for the variables $\mathbf{V}(t)$ and $\delta n(\mathbf{r},t)$ constitute our non-contracted description. The contraction theorem projects this description onto a contracted description involving only one of these two variables. For our purpose this essentially means solving the second equation for $\delta n(\mathbf{r},t)$ and substituting the result in the first equation, leading to a single stochastic equation for the variable  $\mathbf{V}(t)$, which reads
\begin{equation}
M{\frac{d{\bf V}(t)}{dt}}=  -\zeta ^{(s)}{\bf V}(t)+{\bf f}
 ^{(s)}(t)- \int_0^t dt' \Delta
\zeta(t-t')  {\bf V}(t')+ {\bf F} (t), \label{gletracerdif}
\end{equation}
where $M$ is the mass and ${\bf V}(t)$ the velocity of the tracer particle, while $\zeta^{(s)}$ is the (short-time  \cite{pusey,prlhi}) friction coefficient caused by the frictional resistance of the supporting solvent and ${\bf f}^{(s)}(t)$ the associated random force. The memory term involving the time-dependent friction function $\Delta \zeta(t)$, and its associated random force ${\bf F}(t)$, are the friction and fluctuating forces that originate in the time-evolution of the cage formed by the surrounding colloidal particles. Under well defined approximations, the exact result for the time-dependent friction function $\Delta \zeta (t)$ derived in Ref. \cite{faraday} was shown there to reduce to the following approximate expression in terms of the collective and self intermediate scattering functions (ISFs) $F(k,t)$ and $F_S(k,t)$, 
\begin{equation}
\Delta \zeta^* (t) \equiv \Delta \zeta (t)/\zeta ^{(s)} =\frac{D_0}{3\left( 2\pi \right) ^{3}n}\int d
{\bf k}\left[\frac{ k[S(k)-1]}{S(k)}\right] ^{2}F(k,t)F_S(k,t).
\label{dzdt0}
\end{equation}
In this equation $n$ the number concentration, $S(k)$ the static structure factor of the bulk suspension, and $D_0$ is the short-time self-diffusion coefficient, defined in terms of $\zeta ^{(s)}$ by Einstein's relation,  $D_0 \equiv k_BT/\zeta ^{(s)}$, with   $T$ being  the temperature.

One can also use the GLE method to derive the most general expression for the collective intermediate scattering function $F(k,t)$ of a colloidal dispersion. For this, one starts by considering again the abstract GLE in Eq. (\ref{gle0}), but now with the vector $\delta{\bf a} (t)$ defined as $\delta{\bf a} (t)\equiv \left[ \delta n({\bf k},t),\delta j_{l}({\bf k},t)\right] $ with $\delta n({\bf k},t)$ being the Fourier transform of the fluctuations $\delta n({\bf r} ,t)$ and $\delta j_{l}({\bf k},t)= j_{l}({\bf k},t)\equiv \widehat{{\bf k}}{\bf \cdot j}({\bf k},t)$ being the longitudinal component of the current ${\bf j}({\bf k},t)$. The continuity equation couples $(\partial \delta n({\bf k},t)/\partial t)$ with ${\bf j}({\bf k},t)$ and the format of Eq. (\ref{gle0}) determines the most general time evolution equation of the current.  The two stochastic equations for $\delta n({\bf k},t)$  and $\delta j_{l}({\bf k},t)$ are then written, using the contraction theorem, as a single equation with memory for $\delta n({\bf k},t)$ which, after multiplying by $\delta n(-{\bf k},0)$, becomes a time-evolution equation for the ISF $F(k,t)\equiv <\delta n({\bf k},t)\delta n(-{\bf k},0)>$. This exercise, done in detail in ref. \cite{scgle1}, after some additional refinements and approximations \cite{todos2}, finally leads to the following expression for the  Laplace transform $F(k,z)$ of $F(k,t)$, namely,  
\begin{equation}
F(k,z)=\frac{S(k)}{z+\frac{k^{2}D_{0}S^{-1}(k)}{1+ \lambda (k) \Delta \zeta^*
(z)}},
\label{fkz}
\end{equation}
with $\lambda (k)\equiv [1+(k/k_c)^2]^{-1}$ being a phenomenological interpolating function, with $k_c$ being the only empirically-determined parameter \cite{todos2}. Similar arguments lead to an analogous approximate result for the \emph{self} ISF  $F_S(k,t)$,
\begin{equation}
F_{S}(k,z)=\frac{1}{z+\frac{k^{2}D_{0}}{1+ \lambda (k) \Delta \zeta^*(z)}}. 
\label{fskz}
\end{equation}

Eqs. (\ref{dzdt0}), (\ref{fkz}), and  (\ref{fskz}) constitute a closed system of equations for the dynamic properties 
$F(k,t)$, $F_{S}(k,t)$, and $\Delta \zeta^*(z)$ for given $S(k)$. These equations constitute the core of the  theory of the dynamic properties of liquids \emph{in thermodynamic equilibrium}, referred to as the  \emph{Self-Consistent Generalized Langevin Equation} (SCGLE) theory, first  presented in the year 2000  \cite{tesislaura}, and further elaborated and applied in subsequent scientific publications \cite{scgle1,scgle2,scgle3,scgle4}. This theory soon demonstrated  to yield highly accurate predictions of the dynamics of a wide range of model liquids, including hard and soft spheres with or without attractions, mixtures of spherical particles with or without electric charges, etc. Here we shall not describe these advances, since they were reviewed in detail in 2010 in Ref. \cite{reviewroque}. We do highlight, however, the fact that the SCGLE theory  predicts the existence of a region of the state space  (low temperatures and/or high densities) where the metastable fluid  might be unable to reach its thermodynamic equilibrium state, becoming trapped in glassy macroscopic states \cite{rmf,todos1,todos2}. In this regard, the SCGLE theory becomes equivalent in many respects to the mode coupling theory of the glass transition \cite{goetze1,goetze2,goetze3,goetze4,mctjenssen}.

\subsection{Summary. }\label{subsection2.5}

In summary, it is true that the main  limitations of the Onsager-Machlup theory of thermal fluctuations  (being Gaussian and  Markov) are imposed by the underlying mathematical model, the  Ornstein-Uhlenbeck stochastic process. Hence, the route to ``liberate'' this theory from such limitations was to choose a more flexible mathematical model. For this, the stationarity theorem \cite{delrio} became the fundamental cornerstone that allowed the definition of a  simple mathematical model of non-Markov and not-necessarily-Gaussian stochastic processes whose only essential condition is stationarity. In this manner, removing these limitations of the Onsager-Machlup theory became a mere technical issue, which allowed us to recognize the GLE in Eq. (\ref{gle0}) as an essential consequence of the mathematical condition of stationarity. In this section we also illustrated the concrete application of the resulting GLE formalism with the derivation of the exact fundamental equations for the main dynamic properties, which led to the approximate SCGLE theory of the dynamics of equilibrium liquids.

\section{Piecewise-stationary (PS) stochastic process.}\label{section3}

The major limitation of the Onsager-Machlup theory is its \emph{physical} restriction to the description of thermal fluctuations around \emph{thermodynamic equilibrium} states. As recently explained in Ref.  \cite{OUgranular}, however, there is no fundamental impediment to use the Ornstein-Uhlenbeck stochastic process to describe fluctuations also in \emph{non-equilibrium stationary} systems. This indicates a route to extend the Onsager-Machlup theory, including its non-Gaussian and non-Markov versions, to systems that are not in thermodynamic equilibrium and are not even stationary. The corresponding discussion is the subject of this section, in which we introduce a simple mathematical model that extends the Ornstein-Uhlenbeck stochastic process to non-stationary conditions. Such an extended model is referred to as  \emph{piecewise-stationary (PS) stochastic process}, which will be employed in the following section to develop the non-equilibrium extension of the Onsager-Machlup theory.

\subsection{Formal definition of the PS stochastic process.}\label{subsection3.1}

A stochastic process $\mathbf{a} (t)=(a_1(t),a_1(t),...., a_\nu(t))^T$   is formally defined in terms of the joint probability density $W_m(\textbf{a}_1,t_1;\textbf{a}_2,t_2;...;\textbf{a}_m,t_m)$ for the state vector $\textbf{a}(t)$ to have a value in the interval $\textbf{a}_i \le \textbf{a}(t_i) \le \textbf{a}_i + d \textbf{a}_i$ for $i=1,2,...,m$, for all possible positive integer values of $m$ and all possible sets of times $(t_1,t_2,...,t_m)$. If we wish to describe non-stationary conditions, none of these probability densities, or the properties that derive from them, can be assumed to be time-translational invariant. For example, the mean value
\begin{equation}
\overline{ {\textbf a }(t)} = \int d \mathbf{a}_1  \ \mathbf{a}_1 \ W_1(\textbf{a}_1,t),
\label{nonstat1stmoment}
\end{equation}
 the covariance
\begin{equation}
{\bm{\sigma}}(t)\equiv \int d \mathbf{a}_1  \ [\mathbf{a}_1-\overline{\mathbf{a}(t)}][\mathbf{a}_1-\overline{\mathbf{a}(t)}]^T \ W_1(\textbf{a}_1,t),
\label{nonstatcovariance}
\end{equation}
and all the moments of $W_1(\textbf{a},t)$, do depend on $t$. Similarly, the 2-time correlation  function
\begin{equation}
\mathbf{C}(t,t')\equiv \int d \mathbf{a}_1 \int d \mathbf{a}_2 \ [\mathbf{a}_1-\overline{\mathbf{a}(t)}][\mathbf{a}_2-\overline{\mathbf{a}(t')}]^T \ W_2(\textbf{a}_1,t;\textbf{a}_2,t')
\label{nonstat2tcf}
\end{equation}
depends on both times, $t$ and $t'$. Determining all the moments of all the probability densities $W_m(\textbf{a}_1,t_1;\textbf{a}_2,t_2;...;\textbf{a}_m,t_m)$ is, of course, impractical and perhaps impossible in general. Thus, we must choose some simplifying strategy to identify less general but non-trivial and tractable models of non-stationary processes.

For this, rather than attempting to state our simplifying assumptions at the level of all the probability densities $W_m(\textbf{a} _1,t_1;\textbf{a}_2,t_2;...;\textbf{a}_m,t_m)$, we shall only focus \emph{on the moments} of $W_1(\textbf{a}_1,t)$ and $W_2(\textbf{a}_1,t;\textbf{a}_2,t')$, and in fact, only on the most immediate and  relevant ones, namely, ${\overline {\textbf a }}(t)$, ${\bm{\sigma}}(t)$, and $\mathbf{C}(t,t')$. In fact, we now define the PS stochastic process by means of two basic defining assumptions, from which these moments can be derived. 

\vskip1cm
\noindent {\bf Definition of the piecewise-stationary (PS) stochastic process.} A piecewise-stationary stochastic process $\mathbf{a} (t)=(a_1(t),a_1(t),...., a_\nu(t))^T$  is a globally non-stationary but locally stationary stochastic process, defined by one basic assumption imposed on its conditioned mean value ${\overline {\textbf a }}^0(t)$ and another on the fluctuations  $\delta {\textbf a}(t+\tau) \equiv {\textbf a }(t+\tau)-{\overline {\textbf a }}^0(t)$ around ${\overline {\textbf a }}^0(t)$  which, at fixed evolution time $t$, and as a function of the fluctuation time $\tau$,  behave as a general stationary non-Markov process. 


\vskip1cm
\noindent $\bm{First\ Basic\ Assumption}$:  The first moment ${\overline {\textbf a }}^0(t)$ will be provided by  the solution  of a deterministic non-linear differential equation of  the general form
\begin{equation}
\frac{d{\overline{\textbf a}}(t)}{dt}= \mathcal{R}\left[{\overline {\textbf a }}(t)   \right],
\label{meanvalue0}
\end{equation}
whose  possible stationary solutions (which solve $\mathcal{R}\left[{\textbf a }  \right]=0$) must include at least one ``atractor''. This  is a  \emph{stable} stationary solution, denoted as ${\textbf a }^{ss}$,  such that any solution ${\overline{\textbf a}}^0(t)$ of Eq. (\ref{meanvalue0}), with initial condition ${\textbf a}^0={\overline{\textbf a}}^0(t=0)$ in the basin of attraction of ${\textbf a }^{ss}$, is \emph{asymptotically stable}. This implies the existence of a Lyapunov function $\Lambda [{\textbf a }]$,  such that $\Lambda(t)\equiv \Lambda[\overline{\textbf a}(t)]$  is a positive decreasing function of $t$ when $\overline{\textbf a}(t)\to {\textbf a }^{ss}$ \cite{keizer}.


\vskip1cm
\noindent $\bm{Second\ Basic\ Assumption}$:  The stochastic process ${\textbf a }(t)$ involves two distinct time scales, such that for any fixed time $t$, the fluctuations $\delta {\textbf a}(t+\tau) \equiv {\textbf a }(t+\tau)-{\overline {\textbf a }}^0(t)$ around the conditioned mean value ${\overline {\textbf a }}^0(t)$ can be approximated by a non-Markov stationary stochastic process, defined by the solutions of a linear stochastic equation  of the general kind
\begin{equation}
\frac{d \delta {\textbf a}(t+\tau)}{d\tau}=   \int_0^\tau {\textbf H}(\tau-\tau';t) \cdot \delta {\textbf a}(t+\tau') d\tau' +{\textbf f}(t+\tau),
\label{nonmarkovlangevineq1}
\end{equation}
with the additive noise ${\textbf f }(t+\tau)$ being, for fixed $t$, not  necessarily Gaussian nor $\delta$-correlated, but necessarily stationary,  with zero mean  $\overline{{\textbf f}(t+\tau)}={\textbf 0}$, correlation function 
\begin{equation}
\overline{{\bf f}(t+\tau){\bf f}^{T}(t+\tau')}=  {\bm{\Gamma}}(\tau-\tau';t),
\label{noisecorrelfunction}
\end{equation} 
and uncorrelated with the initial condition $\delta {\textbf a }(t+0)$, i.e.,  $\overline{{\textbf f}(t+\tau)\delta {\textbf a }^{T}(t+0)}=0$.

\subsection{Qualitative notion of the PS stochastic process.}\label{subsection3.2}

\begin{figure*}[ht]
\includegraphics[scale=.4]{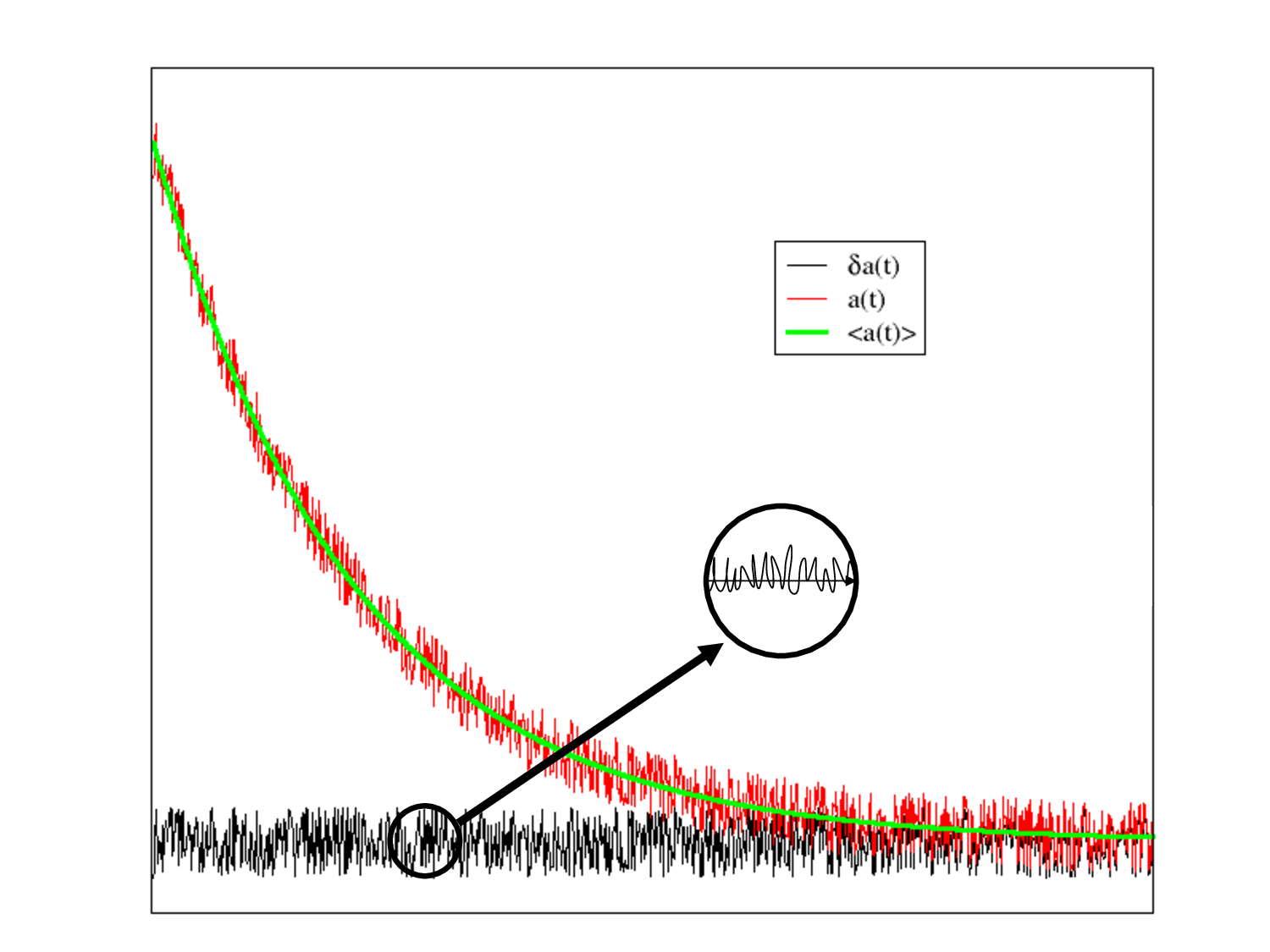}
\caption{Schematic illustration of one realization of the non-stationary stochastic process $ {\textbf a }(t)$ (red noisy curve), which fluctuates around the mean value ${\overline {\textbf a }}(t)$ (green solid curve), with the difference $\delta {\textbf a}(t)\equiv  {\textbf a}(t) - {\overline {\textbf a }}(t)$ represented by the black noisy curve.}
\label{fig1}
\end{figure*}

The main qualitative feature of the piecewise stationary stochastic process  ${\textbf a }(t)$, formally stated by the second basic assumption above, is schematically illustrated in Fig. \ref{fig1}, which sketches a typical realization of $ {\textbf a }(t)$ (red noisy curve) that fluctuates rather fast around the much slower mean value ${\overline {\textbf a }}(t)$ (green solid curve). Thus,  although  ${\textbf a }(t)$ is manifestly non-stationary, the difference $\delta {\textbf a}(t)\equiv  {\textbf a}(t) - {\overline {\textbf a }}(t)$ (black noisy curve) has zero mean, and its illustrative realization  in the figure seems somewhat stationary. In reality, however,  the corresponding (conditioned) covariance ${\bm{\sigma}}(t)\equiv \overline{ \delta {\textbf a}(t) \delta {\textbf a}^T(t)}$ is also time-dependent.

The qualitative idea is that, although the process is  globally non-stationary, we expect the time-evolution of ${\overline {\textbf a }}(t)$ and ${\bm{\sigma}}(t)$ to be much slower than the dynamics of the fluctuations during the short time interval  $t_n\le t \le t_{n+1}$, illustrated by the enlarged circled portion of the figure. We might then model the stochastic process $\delta {\textbf a}(t)$, now written as  $\delta {\textbf a}(t_n+\tau)\equiv {\textbf a }(t_n+\tau)-{\overline {\textbf a }}(t_n)$, as approximately stationary, i.e., within this short time interval, it can be described by the general stationary non-Markov process described in the previous section. This local stationarity is the defining element of our non-stationary stochastic model referred to as \emph{piecewise-stationary (PS)} stochastic process, whose most precise definition \cite{nescgle0} is summarized by the second basic assumption above.

\subsection{General properties of the PS stochastic process.}\label{subsection3.3}

From the $\bm{First\ and\ Second\ Basic\ Assumptions}$ above one can derive a number of consequences, and here we list only a few.
\vskip0.5cm

1. One manner to explicitly guarantee that  Eq. (\ref{meanvalue0}) has at least one  \emph{stable} stationary solution ${\textbf a }^{ss}$, as required by our $\bm{First\ Basic\ Assumption}$,  is to factorize the function $\mathcal{R}\left[ {\textbf a }   \right]$ as  $\mathcal{R}\left[ {\textbf a }  \right]=\mathcal{L}\left[ {\textbf a }   \right]\cdot \Delta {\textbf F}[ {\textbf a } ]$, with $\Delta {\textbf F}[{\textbf a }] \equiv {\textbf F}[{\textbf a }] - {\textbf F}[{\textbf a }^{ss}]$, and with the vector function ${\textbf F}\left[ {\textbf a }\right]$ chosen as 
\begin{equation}
{\textbf F}\left[ {\textbf a }\right]\equiv \left( \frac{\partial \Lambda [ {\textbf a }]}{ \partial {\textbf a }}\right),
\label{efesdex}
\end{equation}
with $\Lambda [ {\textbf a}]$ being the Lyapunov function associated with the stability of the stationary solution ${\textbf a }^{ss}$. 

2. This allows us to rewrite the time-evolution equation of ${\overline{\textbf a}}(t)$ in  Eq. (\ref{meanvalue0}) as 
\begin{equation}
\frac{d{\overline{\textbf a}}(t)}{dt}= \mathcal{L}\left[{\overline {\textbf a }}(t)   \right]\cdot \Delta {\textbf F}[{\overline {\textbf a }}(t)]. 
\label{meanvalue1}
\end{equation}

3. In the long-$t$ regime, where  the deviations $\Delta {\overline{\textbf a}}(t)\equiv {\overline{\textbf a}}(t)- {\textbf a }^{ss}$ from stationarity are already small, we can  approximate $\Delta {\textbf F}[{\overline{\textbf a}}(t)]$ to linear order in the deviations as $\Delta {\textbf F}[{\overline{\textbf a}}(t)] \approx  \mathcal{E}[ {\textbf a }^{ss}]   \cdot \Delta {\overline {\textbf a }}(t)$, with the stability matrix function $ \mathcal{E}[{\textbf a}]$ defined as
\begin{equation}
 \mathcal{E}[{\textbf a }]\equiv \left( \frac{\partial {\textbf F} [{\textbf a }]}{\partial {\textbf a}}      \right) =\left( \frac{\partial ^2 \Lambda [{\textbf a }]}{\partial {\textbf a}\  \partial {\textbf a }}      \right).
\label{vquad}
\end{equation}
\bigskip
This leads to the linearized version of Eq. (\ref{meanvalue1}), namely,
\begin{equation}
\frac{d{\overline{\textbf a}}(t)}{dt}= \mathcal{L}\left[{\textbf a }^{ss}   \right]\cdot  \mathcal{E}[ {\textbf a }^{ss}]   \cdot \Delta {\overline {\textbf a }}(t).
\label{meanvalue2}
\end{equation}

\vskip1cm
On the other hand, from the $\bm{Second\ Basic\ Assumption}$, some general properties of the PS stochastic process follow:

\medskip 
4. The most relevant is that the local (or ``piecewise'')   application of the theorem of stationarity \cite{delrio} implies that Eq.  (\ref{nonmarkovlangevineq1}) must have the  rigid mathematical structure of  Eq.  (\ref{gle0}), which adapted to our two-time-scales notation, is described by the following $t$-dependent  generalized Langevin equation ($t$-GLE),
\begin{equation}
\frac{d\delta  {\textbf a }(t+\tau)}{d \tau}= - {\bm{\omega}} (t) \cdot {\bm{\sigma}}^{-1}(t)
\cdot \delta  {\textbf a }(t+\tau) - \int_0^{\tau}d\tau'  {\bm{\Gamma}}(\tau-\tau';t)\cdot
{\bm{\sigma}}^{-1}(t)\cdot \delta {\textbf a }(t+\tau')+{\textbf f }(t+\tau).
\label{glen0}
\end{equation}
In this equation ${\bm{\omega}}(t)$ is an antisymmetric matrix, ${\bm{\omega}}^T(t)=-{\bm{\omega}}(t)$, and the matrix  $ {\bm{\Gamma}}(\tau-\tau';t)$ is the $\tau$-dependent correlation function of the additive noise,
${\bm{\Gamma}}(\tau-\tau';t) =  {\bm{\Gamma}}^{T}(\tau'-\tau;t)=\overline{{\bf f}(t+\tau){\bf f}^{T}(t+\tau')}$.  This implies that the memory matrix $ {\textbf H}(\tau-\tau';t)$ of Eq. (\ref{nonmarkovlangevineq1}) can be written as 
\begin{equation}
{\textbf H}(\tau;t)= - [ {\bm{\omega}}(t) 2\delta (\tau)+ {\bm{\Gamma}}(\tau;t)]  \cdot {\bm{\sigma}}^{-1}(t).  
\label{hdetaut}
\end{equation}

\medskip
5. In the Markov limit, ${\textbf H}(\tau-\tau';t)=   \mathcal{H}(t) 2\delta (\tau-\tau')$, with
\begin{equation}
\mathcal{H}(t)  \equiv \int_0^\infty {\textbf H}(\tau;t) d\tau = - [ {\bm{\omega}}(t) + {\bm{\gamma}}(t)]  \cdot {\bm{\sigma}}^{-1}(t),
\label{hmarkov3}
\end{equation}
with
\begin{equation}
{\bm{\gamma}}(t)  \equiv \int_0^\infty  {\bm{\Gamma}}(\tau;t)d\tau = \int_0^\infty  \overline{{\bf f}(t+\tau){\bf f}^{T}(t)}  d\tau.
\label{gammarkov2}
\end{equation}
This general result provides the mathematical basis to extend the well-known equilibrium Green-Kubo relations \cite{hansen}  to non-equilibrium conditions. 

6. In the same limit, the non-Markov stochastic equation (\ref{nonmarkovlangevineq1}) becomes \begin{equation}
\frac{d\delta  {\textbf a }(t+\tau)}{d \tau}=\mathcal{H}(t) \cdot \delta  {\textbf a }(t+\tau)+{\textbf f }(t+\tau),
\label{glen22}
\end{equation}
where ${\textbf f }(t+\tau)$ is, for fixed $t$, a stationary $\delta$-correlated noise of zero mean and correlation function  ${\overline {{\textbf f }(t+\tau){\textbf f }^{\ T}(t+\tau')}} = 2 \delta(\tau-\tau') {\bm{\gamma}}(t)$.  Thus, locally in time (i.e., as a function of $\tau$ for given  fixed $t$), the Markov limit of our general model of PS stochastic process must satisfy all the properties of the Ornstein-Uhlenbeck process. 
\medskip

7. In particular,  the time-evolution equation of the corresponding non-equilibrium covariance  ${\bm{\sigma}}(\tau;t)\equiv \overline{ \delta {\textbf a}(t+\tau) \delta {\textbf a}^T(t+\tau)}$ can be derived from Eq. (\ref{glen22}) in the same manner as in the genuine stationary (i.e., Ornstein-Uhlenbeck) process, with the following result  \cite{keizer}
\begin{equation}
\frac{d{\bm{\sigma}}(\tau;t)}{d\tau} = \mathcal{H}(t) \cdot  {\bm{\sigma}}(\tau;t) + {\bm{\sigma}}(\tau;t)
\cdot  \mathcal{H}^{ T} (t)+ 2\gamma(t).
\label{dsigmadtirrevlocal}
\end{equation}
In its stationary limit $\tau \to \infty$, this equation implies the following \emph{local} non-equilibrium fluctuation-dissipation relation (FDR),
\begin{equation}
\mathcal{H}(t)\cdot {\bm{\sigma}}^{ss}(t) + {\bm{\sigma}}^{ss}(t) \cdot
\mathcal{H}^{T }(t) +2{\bm{\gamma}}(t) =0, 
\label{fdrelatlocal}
\end{equation}
where ${\bm{\sigma}}^{ss}(t) \equiv {\bm{\sigma}}(\tau \to \infty;t)$.


8. This FDR allows us to rewrite Eq. (\ref{dsigmadtirrevlocal}) as 
\begin{equation}
\frac{d{\bm{\sigma}}(\tau;t)}{d\tau} = \mathcal{H}(t)\cdot
[{\bm{\sigma}}(\tau;t)-{\bm{\sigma}}^{ss}(t)] + [{\bm{\sigma}}(\tau;t)-{\bm{\sigma}}^{ss}(t)] \cdot \mathcal{H}^{T }(t).
\label{sigmadteq2local}
\end{equation}
Recalling that ${\bm{\sigma}}(\tau;t)\equiv {\bm{\sigma}}(t+\tau)$, and  in the limit $\tau\to 0$, this equation leads to the time-evolution equation for the covariance ${\bm{\sigma}}(t)$ in the slower time-scale $t$, namely,
\begin{equation}
\frac{d{\bm{\sigma}}(t)}{dt} = \mathcal{H}(t)\cdot
[{\bm{\sigma}}(t)-{\bm{\sigma}}^{ss}(t)] + [{\bm{\sigma}}(t)-{\bm{\sigma}}^{ss}(t)] \cdot \mathcal{H}^{T }(t).
\label{sigmadteq5}
\end{equation}

9. The fact that the  conditioned mean value $\overline{{\textbf a }(t+\tau)}^0$ has a  transiently stable long-$\tau$ ``stationary'' limit $\overline{{\textbf a }(t)}$, implies that  when $\overline{{\textbf a }(t+\tau)}^0$ approaches $\overline{\textbf a}(t)$ at long $\tau$, the Lyapunov function $\Lambda(t+ \tau)$ approaches  
\begin{equation}
\Lambda(t+ \tau) \equiv  {(\overline{{\textbf a }(t+\tau)}^0-\overline{\textbf a}(t)) }^T \cdot \mathcal{E}(t) \cdot {(\overline{{\textbf a }(t+\tau)}^0-\overline{\textbf a}(t)) }/2,
\label{lambdadxsmalldt}
\end{equation}
where the matrix $\mathcal{E}(t)$ is symmetric and positive definite. It also implies that a  symmetric and positive definite matrix $G(t)$ exists such that
\begin{equation}
\mathcal{H}(t) \cdot \mathcal{E}^{-1}(t) + \mathcal{E}^{-1}(t)\cdot \mathcal{H}^T(t) + G(t)=0.
\label{hvvhtg2dt}
\end{equation}

\medskip
10. Comparing the latter equation with  the fluctuation-dissipation relation in Eq. (\ref{fdrelatlocal}) we immediately identify $G(t)$ with $2{\bm{\gamma}}(t)$ (defined in Eq. (\ref{gammarkov2})), and $\mathcal{E}^{-1}(t)$ with ${\bm{\sigma}}^{ss} (t)$, i.e., 
\begin{equation}
 {\bm{\sigma}}^{ss}(t) \cdot  \mathcal{E}(t) = {\bm I},
\label{sigmaxvsdt}
\end{equation}
where  $\mathcal{E}(t)$ is defined as $\mathcal{E}(t)\equiv \mathcal{E}[\overline{\textbf a}(t)]$, with the stability matrix function $ \mathcal{E}[{\textbf a}]$ defined in Eq. (\ref{vquad}).

\medskip
11. From the $t$-dependent GLE in Eq. (\ref{glen0}) many properties of the PS stochastic process can be derived \cite{nescgle0,nescgle1}. For example, multiplying the right side of this equation by $\delta {\textbf a }^T(t)$ and taking the average, one can derive the time-evolution equation for the locally-stationary time-correlation function ${\textbf C}(\tau;t) \equiv \overline{ \delta {\textbf a }(t +\tau)\delta {\textbf a }^T(t)}$, whose solution can be written as $\hat{{\textbf C}}(z;t) = \left[ z {\textbf I} + \left( {\bm{\omega}}(t) +\hat{ {\bm{\Gamma}}}(z;t) \right) \cdot {\bm{\sigma}}^{-1}(t)      \right]^{-1}\cdot {\bm{\sigma}}(t)$. In this equation $\hat{{\textbf C}}(z;t)$ and $\hat{{\bm{\Gamma}}}(z;t)$ are the Laplace transforms of ${\textbf C}(\tau;t)$ and ${\bm{\Gamma}}(\tau;t)$, and ${\bm{\sigma}}(t) ={\textbf C}(\tau=0;t)$ is the $t$-dependent covariance.

\section{The non-equilibrium Onsager-Machlup theory.}\label{section4}

In this section we will employ the mathematical infrastructure represented by the piecewise stationary stochastic process,  developed in the previous subsection, to propose a  non-equilibrium non-stationary version of the Onsager-Machlup theory in an extremely economic logical fashion. We start by  stating the  two primary physical postulates that constitute the fundamental basis of this theory, and then briefly describe the main implications of these two postulates.


\subsection{Fundamental postulates of the non-equilibrium Onsager-Machlup theory.}\label{subsection4.1}

\bigskip
\noindent{\bf Postulate I}: The statistical properties of the fluctuations  of the extensive thermodynamic variables $\mathbf{a} (t)=(a_1(t),a_1(t),...., a_\nu(t))^T$  of an arbitrary non-equilibrium macroscopic system, can be described by the mathematical model referred to as \emph{piecewise stationary (PS) stochastic process}.

\vskip1cm
\noindent{\bf Postulate II}: A Lyapunov function $\Lambda[{\textbf a}]$ exists that guarantees that  the mean value $\overline{\mathbf{a} }^{eq}$ of each thermodynamic equilibrium state is also a \emph{stable} stationary solution of the deterministic non-linear equation satisfied by the time-dependent mean value $\overline{\mathbf{a}} (t)$. This Lyapunov function $\Lambda[{\textbf a}]$  is postulated to be 
\begin{equation}
\Lambda [{\textbf a }] = - S[{\textbf a }]/k_B, 
\label{equilcondlyapunov}
\end{equation}
where the function $S[{\textbf a }]$ is the thermodynamic  entropy of the system and with $k_B$ being Boltzmann's constant.

\vskip1cm

\subsection{Summary of the non-equilibrium Onsager-Machlup theory.}\label{subsection4.2}

Listing the most salient consequences of these two fundamental Postulates, and discussing their physical meaning, provides a concise summary of our proposed non-equilibrium extension of the Onsager-Machlup theory of thermal fluctuations. To simplify the task of providing such a concise summary, let us for a moment exclude from these two Postulates the identification in Eq. (\ref{equilcondlyapunov}) of the  Lyapunov function $\Lambda[{\textbf a}]$ with the negative entropy. Then, what remains from these two postulates is essentially equivalent to state that  the extensive thermodynamic variables $\mathbf{a} (t)=[a_1(t),a_1(t),...., a_\nu(t)]^T$  of an arbitrary non-equilibrium macroscopic system satisfy the $\bm{First}$ and $\bm{Second\ Basic\ Assumptions}$ that  define the piecewise-stationary (PS) stochastic process in Subsection \ref{subsection3.1}. 

This implies that the properties of the PS stochastic process listed in Subsection \ref{subsection3.3} of the previous section, which derive from those two basic assumptions, already constitute the desired ``concise summary of our proposed non-equilibrium extension of the Onsager-Machlup theory of thermal fluctuations''. Thus, we suggest a second reading of Subsection \ref{subsection3.3}, but now including the identification of the  Lyapunov function $\Lambda[{\textbf a}]$ with the negative entropy, according to Eq. (\ref{equilcondlyapunov}). In this second reading, the properties of the PS stochastic process listed there as mere mathematical consequences of well-defined mathematical assumptions, now acquire a distinct physical meaning. As we now discuss, near the stationary equilibrium conditions, these derived properties turn onto well-known concepts characteristic of the Onsager-Machlup theory of fluctuations and the Onsager theory of irreversible thermodynamics. The simplest practical summary of the \emph{kinetic }aspects of the non-equilibrium Onsager-machlup theory is provided by the time-evolution equations of the most relevant physical properties. 

\subsubsection{Non-linear ``Onsager linear laws'' of irreversible thermodynamics.}\label{subsubsection4.2.1}

The first relevant time-evolution equation is Eq. (\ref{meanvalue0}) for the mean value $\overline{\textbf{a}}(t)$, which is the essence of the first basic assumption of a PS stochastic process. This assumption also imposes the condition of thermodynamic stability  in the kinetics of approach to equilibrium. A sufficient condition for this equation to also comply with this condition, is to write it as suggested by Eq. (\ref{meanvalue1}), i.e., as 
\begin{equation}
\frac{d{\overline{\textbf a}}(t)}{dt}= \mathcal{L}\left[{\overline {\textbf a }}(t)   \right]\cdot \Delta {\textbf F}[{\overline {\textbf a }}(t)]. 
\label{meanvalue1p}
\end{equation}
This is immediately identified with the most general  non-equilibrium version of Onsager linear laws, which state that the ``thermodynamic fluxes'', grouped in the vector $[d{\overline{\textbf a}}(t)/dt]$, are formally proportional to the ``thermodynamic forces'' (or ``affinities'') grouped in the vector  $\Delta {\textbf F}[{\overline {\textbf a }}(t) ] \equiv {\textbf F}[{\overline {\textbf a }}(t) ] - {\textbf F}[{\textbf a }^{eq}]$, now involving the conjugate thermodynamic intensive parameters ${\textbf F}\left[ {\textbf a }\right]\equiv (\partial [-S [ {\textbf a }]/k_B]/ \partial {\textbf a })$  \cite{keizer}. 

Near a thermodynamic equilibrium stationary state, however,  the non-linear equation in Eq.  (\ref{meanvalue1p}) can be linearized as in Eq. (\ref{meanvalue2}), but with $ {\textbf a }^{ss}$ now identified with ${\textbf a }^{eq}$, i.e., as 
\begin{equation}
\frac{d{\overline{\textbf a}}(t)}{dt}= \mathcal{L}\left[{\textbf a }^{eq}   \right]\cdot  \mathcal{E}[ {\textbf a }^{eq}]   \cdot \Delta {\overline {\textbf a }}(t).
\label{meanvalue2p}
\end{equation}
These are the genuine linear transport equations, valid only in the so-called linear regime  \cite{keizer}. Their derivation above exhibits a sort of ``complementarity principle'', in the sense that from the general non-equilibrium theory one recovers in the linear regime the usual linear laws of Onsager's linear  irreversible thermodynamics.

In general, however, as an equation for ${\overline {\textbf a }}(t)$,  Eq. (\ref{meanvalue1}) is still highly non-linear, given the state dependence of the vector  ${\textbf F}[{\overline {\textbf a }}(t) ]$ and of the matrix $\mathcal{L}\left[{\overline {\textbf a }}(t)\right]$. The \emph{thermodynamic equations of state}, represented by the dependence ${\textbf F}={\textbf F}\left[ {\textbf a }\right]$ of the intensive parameters on the state variables ${\textbf a}$, are determined by the so-called \cite{callen}  \emph{fundamental thermodynamic relation} $S=S\left[ {\textbf a }\right]$. This relation, in its turn, can in principle be determined \cite{mcquarrie} using Boltzmann's fundamental principle of statistical mechanics, $S\left[ {\textbf a }\right]=k_B \ln W\left[ {\textbf a }\right]$. In contrast, the state dependence of the \emph{kinetic matrix} $\mathcal{L}\left[ {\textbf a }\right]$, constitute what we can refer to as the  \emph{kinetic equations of state}, which are in general unknown. Identifying the fundamental principles that allow their determination constitutes a permanent challenge of non-equilibrium statistical mechanics. Here we illustrate how this fundamental challenge can be addressed in the specific context of the liquid state, which is precisely one of the objectives of the approximate theories referred to as the non-equilibrium self-consistent generalized Langevin equation (NE-SCGLE) theory, described in the following section.

\subsubsection{Time-evolution equation for the non-equilibrium covariance ${\bm{\sigma}} (t)$.}\label{subsubsection4.2.2}

The second observable of particular relevance is the non-equilibrium covariance ${\bm{\sigma}} (t)\equiv \overline{\delta {\textbf a }(t)\delta {\textbf a }^{ \dagger}(t)}$, whose time-evolution equation is Eq. (\ref{sigmadteq5}), with  the $t$-dependent  ``instantaneous''  \emph{stationary} covariance ${\bm{\sigma}}^{ss}(t)$ given by  $ {\bm{\sigma}}^{ss}(t)  =\mathcal{E}^{-1}(t)$ (see Eq. (\ref{sigmaxvsdt})). The end result is, thus,  
\begin{equation}
\frac{d{\bm{\sigma}}(t)}{dt} = \mathcal{H}(t)\cdot
[{\bm{\sigma}}(t)-\mathcal{E}^{-1}(t)] + [{\bm{\sigma}}(t)-\mathcal{E}^{-1}(t)] \cdot \mathcal{H}^{T }(t),
\label{sigmadteq5p}
\end{equation}
with the $t$-dependent thermodynamic stability matrix $\mathcal{E}(t)$ defined as 
\begin{equation}
\mathcal{E}_{ij}(t) \equiv \mathcal{E}_{ij}[\overline{\textbf a}(t)]  = \left(\frac{\partial F_i[{\textbf a }]}{\partial a_j} \right)_{{\textbf a}=\overline{\textbf a}(t)}=\left(
\frac{\partial^2 \left[ -S[{\textbf a }]/k_B\right]}{\partial a_i\partial a_j} \right)_{{\textbf a}=\overline{\textbf a}(t)},
\label{matrixE2}
\end{equation}
with $S[{\textbf a }]$ being the entropy and $F_i[{\textbf a }] \equiv (\partial [-S[{\textbf a }]/k_B]/\partial a_i)$ the intensive variable conjugate to $a_i$.

Eq. (\ref{sigmadteq5p}) also involves the matrix $\mathcal{H}(t)$, defined by Eq. (\ref{hmarkov3}). According to  Eqs. (\ref{nonmarkovlangevineq1}) and (\ref{glen22}), however, it can also be written as 
\begin{equation}
\mathcal{H}(t)= - [ {\bm{\omega}}(t) + {\bm{\gamma}}(t)]  \cdot {\bm{\sigma}}^{ss-1}(t),  
\label{nmfluctdisipneom}
\end{equation}
with  ${\bm{\omega}}(t)$ being an antisymmetric matrix ${\bm{\omega}}(t)$ representing non-dissipative  mechanical couplings or transport (e.g., convective)  mechanisms, whereas ${\bm{\gamma}}(t)$ originates from the genuine  dissipative processes of transport. According to Eqs.   (\ref{noisecorrelfunction}) and (\ref{gammarkov2}) the matrix  ${\bm{\gamma}}(t)$, referred to as the non-equilibrium matrix of Onsager's kinetic coefficients,  is given by 
\begin{equation}
{\bm{\gamma}}(t) \equiv \int_0^\infty  {\bm{\Gamma}}(\tau;t)d\tau = \int_0^\infty  \overline{{\bf f}(t+\tau){\bf f}^{T}(t)}  d\tau, 
\label{primitivegreenkubo}
\end{equation}
which extends to non-equilibrium the well-known Green-Kubo relations \cite{hansen}. 

Finally, from Property \# 11 listed in Subsection \ref{subsection3.3} we can conclude that the time-correlation function ${\textbf C}(\tau;t) \equiv \overline{ \delta {\textbf a }(t +\tau)\delta {\textbf a }^T(t)}$ can be written as  
\begin{equation}
\hat{{\textbf C}}(z;t) = \left[ z {\textbf I} + \left( {\bm{\omega}}(t) +\hat{ {\bm{\Gamma}}}(z;t) \right) \cdot {\bm{\sigma}}^{-1}(t)      \right]^{-1}\cdot {\bm{\sigma}}(t), 
\label{formalsolcorrfunctftpws}
\end{equation}
where  $\hat{{\textbf C}}(z;t)$ and $\hat{{\bm{\Gamma}}}(z;t)$ are the Laplace transforms of ${\textbf C}(\tau;t)$ and ${\bm{\Gamma}}(\tau;t)$, and ${\bm{\sigma}}(t) ={\textbf C}(\tau=0;t)$ is the $t$-dependent covariance.

\subsubsection{Equilibrium vs. non-equilibrium stationary states.}\label{subsubsection4.2.1}

An important feature to notice is that Postulate I recognizes that all the known empirical transport equations that describe the experimentally measurable temporal evolution of the thermodynamic variables are intrinsically deterministic non-linear differential equations, symbolically represented by Eq. (\ref{meanvalue0}). Postulate II then assumes the existence of thermodynamic equilibrium states, as a particular category of stable stationary solutions of these transport equations. An immediate consequence is that the kinetics of equilibration is  explicitly consistent with  the second law of thermodynamics, since the identification of the Lyapunov function  with $\Lambda (t) = -S[\overline{\textbf{a}}(t)] /k_B$ determines that the time-dependent entropy $S[\overline{\textbf{a}}(t)]$ can only increase with time until  reaching its maximum at  $\overline{\textbf a }(t \to \infty) = {\textbf a }^{eq}$. Thus, the thermodynamic equilibrium stationary solutions must simultaneously satisfy two conditions: to be stationary, $ \mathcal{R}\left[ {\textbf a }^{eq}  \right]={\textbf 0}$, and to maximize the entropy, $(\partial S[\textbf{a}]/ \partial \textbf{a}) _{{\textbf a }={\textbf a }^{eq}}={\textbf 0 }$.

Although the latter may seem a minor observation, it actually bears a deep and fundamental significance. We are quite familiar with thermodynamic equilibrium states, which can be determined solely on the basis of the maximum entropy principle, without appealing to its stationarity condition. The non-linearity of Eq. (\ref{meanvalue0}), however, offers the possibility of additional stationary solutions that do satisfy the condition  $ \mathcal{R}\left[ {\textbf a }^{ss}  \right]={\textbf 0}$, without maximizing the entropy. The physical realization of these \emph{non-equilibrium} stationary solutions could be illustrated, for example, by Prigogine's dissipative structures in open systems \cite{nicolisprigogine}. These non-equilibrium stationary states can be explained, however, on the basis of the local equilibrium approximation, which assumes the existence of only  the fundamental catalog of thermodynamic equilibrium states, i.e., the universal set of solutions of the maximum entropy condition $(\partial S[\textbf{a}]/ \partial \textbf{a}) _{{\textbf a }={\textbf a }^{eq}}={\textbf 0 }$.

However, a far more surprising, but far less studied possibility, refers to non-equilibrium stationary solutions in \emph{closed} systems. As it was demonstrated a little  more than  one  decade ago \cite{nescgle1,nescgle3}, these non-equilibrium stationary solutions  corresponds to the dynamically arrested states in which matter is quite frequently trapped in the form of glasses, gels, etc. This unprecedented revelation (explained in more detail in Section \ref{section5}) is the main high-impact contribution, of the present non-equilibrium generalization of the Onsager-machlup theory.

Let us finally notice that Postulate I above does not assume that the stable stationary state characterized by the mean value ${\textbf a }^{ss}$ (and covariance ${\bm{\sigma}}^{ss}$) correspond to a \emph{thermodynamic equilibrium} state of a thermal system. Hence, its implications and consequences could in principle be tested in systems that satisfy the fundamental assumptions  of that postulate (basically Eq. (\ref{meanvalue0}) for the mean value $\overline{\textbf a}(t)$ and Eq. (\ref{nonmarkovlangevineq1}) for the fluctuations), and that relax to \emph{stationary} but non-equilibrium conditions characterized by the stationary mean value ${\textbf a }^{ss}$  and covariance   ${\bm{\sigma}}^{ss}$. These considerations, however, were never part of the original Onsager-Machlup theory, which starts by restricting itself to  thermal fluctuations in systems at thermodynamic equilibrium. In our present generalization of the Onsager-Machlup theory, instead, we may or may not adopt  the actual link with genuine macroscopic thermal system and their thermodynamic equilibrium states, embodied in the identification of the Lyapunov function with the negative entropy, $- S[{\textbf a }]/k_B$, introduced in Postulate II.


\vskip2cm
\section{General theory of irreversible processes in liquids.}\label{section5}

In this section we briefly describe how the general and abstract formalism of the non-equilibrium Onsager-Machlup theory, presented in the previous section, turns into the generic theory of the non-equilibrium evolution of the structure and dynamics of simple liquids, referred to as the non-equilibrium self-consistent generalized Langevin equation (NE-SCGLE) theory, first presented and explained in full detail in Ref.  \cite{nescgle1}.

\subsection{From a general and abstract formalism to a concrete but generic theory}\label{subsection9.1}

To apply this canonical formalism one must start by defining which physical properties are represented by the components $a_i(t)$ of the abstract state vector ${\textbf a }(t)$. 
For example, having in mind a monocomponent liquid formed by $N$ particles in a volume $V$, we may identify $a_i(t)$ with the instantaneous number $N_i(t)$ of particles in the volume $\Delta V=V/M$ of the \emph{i}th cell of an
(imaginary) partitioning of the volume $V$ into $M$ cells. Or,
better, with the ratio $n_i(t)\equiv N_i(t)/\Delta V$, which
in the limit $\Delta V/V \to 0$ becomes the local particle
concentration profile $n(\textbf{r},t)$. As explained in detail
in Ref. \cite{nescgle1}, with this identification of the state vector ${\textbf a }(t)$, the abstract equations (\ref{meanvalue1p}) for the mean value $\overline{{\textbf a }}(t)$ and (\ref{sigmadteq5p}) for the covariance ${\bm{\sigma}}(t)$, become the concrete but generic (i.e.,
applicable to any monocomponent liquid)  time-evolution equations
for the mean value $\overline{n}(\textbf{r},t)$ and for the
covariance $\sigma(\textbf{r},\textbf{r}';t)\equiv
\overline{\delta n (\textbf{r},t)\delta n (\textbf{r}',t)}$ of
the fluctuation $\delta n (\textbf{r},t) = n (\textbf{r},t)-
\overline{n}(\textbf{r},t)$. The first of these equations reads
\begin{equation} \frac{\partial \overline{n}(\textbf{r},t)}{\partial
t} = D^0{\nabla} \cdot b(\textbf{r},t)\overline{n}(\textbf{r},t)
\nabla \beta\mu[{\bf r};\overline{n}(t)], \label{difeqdlp}
\end{equation}
whereas the second is written in terms of the Fourier transform
(FT) $\sigma(k;\textbf{r},t)$ of the globally non-uniform but
locally homogeneous covariance $ \sigma(\mid \textbf{x}\mid;\textbf{r},t) \equiv \sigma (\textbf{r},\textbf{r} +
\textbf{x};t)$,
\begin{eqnarray}
\begin{split}
\frac{\partial \sigma(k;\textbf{r},t)}{\partial t} = & -2k^2 D^0
\overline{n}(\textbf{r},t) b(\textbf{r},t)
\mathcal{E}(k;\overline{n}(\textbf{r},t)) \sigma(k;\textbf{r},t)
\\ & +2k^2 D^0 \overline{n}(\textbf{r},t)\ b(\textbf{r},t). \label{relsigmadif2p}
\end{split}
\end{eqnarray}
In these equations $D^0$ is the particles' short-time self-diffusion coefficient \cite{pusey} and $b(\textbf{r},t)$ is their local reduced mobility. In addition, $\beta\mu[{\bf r};\overline{n}(t)]$ is the chemical potential per particle (in units of the thermal energy $k_BT=\beta^{-1}$),  determined by  $\beta\mu[{\bf r};\overline{n}(t)] \equiv- k^{-1}_B(\delta S[n;T]/\delta n(\textbf{r}))_{n=\overline{n}(t)}$, where $S[n;T]$ is the entropy, in general a  \emph{functional} of the density. The second functional derivative of $S[n;T]$ determines the stability matrix $\mathcal{E}[{\bf r},\textbf{r}';n]\equiv - k^{-1}_B(\delta^2 S[n;T]/\delta n(\textbf{r})\delta n(\textbf{r}')) = \left( {\delta \beta\mu [{\bf r};n]}/{\delta n(\textbf{r}')}\right)$. The function $\mathcal{E}(k;\overline{n}(\textbf{r},t))$ that appears in Eq. (\ref{relsigmadif2p})
is the FT of $\mathcal{E}[{\bf r},\textbf{r} + \textbf{x};n]\equiv
\left[ {\delta \beta\mu [{\bf r};n]}/{\delta n(\textbf{r} +
\textbf{x})}\right]$.

Eqs. (\ref{difeqdlp}) and (\ref{relsigmadif2p}) above correspond
to Eqs. (4.1) and (4.3) of Ref. \cite{nescgle1}, which discusses
other more specific theories and limits that turn out to be
contained as particular cases of these equations. For example, let
us imagine that we manipulate the system to an arbitrary
(generally non-equilibrium) initial state with mean concentration
profile $\overline{n}^{0}(\textbf{r})$ and covariance
$\sigma^{0}(k;\textbf{r})$, for then letting the system
equilibrate for $t>0$ in the presence of an external field
$\psi({\bf r})$ and in contact with a temperature bath of
temperature $T$. The solution of Eqs. (\ref{difeqdlp}) and
(\ref{relsigmadif2p})  then describes how the system relaxes to
its final equilibrium state whose mean profile and covariance are
$\overline{n}^{eq}(\textbf{r})$ and $\sigma^{eq}(k;\textbf{r})$.
Describing this response at the level of the mean local
concentration profile $\overline{n}(\textbf{r},t)$ is precisely
the aim of \emph{dynamic} density functional theory (DDFT)
\cite{tarazona1, archer}, whose central equation is
recovered from Eq. (\ref{difeqdlp}) in the limit in which we
neglect the friction effects embodied in $b(\textbf{r},t)$ by
setting $b(\textbf{r},t)=1$ (see Eq. (15) of Ref.
\cite{tarazona1}).

The description of the non-equilibrium state of the system in
terms of the random variable $n(\textbf{r},t)$ is not complete,
however, without the simultaneous description of the relaxation of
the covariance $\sigma(k;\textbf{r},t)$ in Eq.
(\ref{relsigmadif2p}). In fact, under some circumstances, the main
signature of the non-equilibrium evolution of a system may be
embodied not in the temporal evolution of the mean value
$\overline{n}({\bf r};t)$ but in the evolution of the covariance
$\sigma(k;\textbf{r},t)$ (which is essentially a non-uniform and
non-equilibrium version of the static structure factor). This may
be the case, for example, when a homogeneous system in the absence
of external fields remains approximately homogeneous,
$\overline{n}({\bf r};t)\approx \overline{n}\equiv N/V$, after a
sudden temperature change. Under these conditions, the
non-equilibrium process is described only by the solution of Eq.
(\ref{relsigmadif2p}). Let us point out that in the limit $b(t)
\to 1$ and within the small-wave-vector approximation,
$\mathcal{E}(k;\overline{n})\approx \mathcal{E}_0 +\mathcal{E}_2
k^2$,  Eq. (\ref{relsigmadif2p}) becomes the basic kinetic
equation describing the early stage of spinodal decomposition
(see, for example, Eq. (3.4) of Ref. \cite{furukawa}).

Eqs. (\ref{difeqdlp}) and (\ref{relsigmadif2p}) above are coupled
between them through the local mobility function
$b(\textbf{r},t)$, essentially a non-stationary and
state-dependent Onsager's kinetic coefficient. In addition, these
two equations are also coupled, through  $b(\textbf{r},t)$, with
the two-point (van Hove) correlation function $C(\textbf{r},\tau;
\textbf{x};t)\equiv \overline{\delta n(\textbf{x},t)\delta n
(\textbf{x}+\textbf{r},t+\tau)}$. According to Ref.
\cite{nescgle1}, the memory function of $C(\textbf{r},\tau;
\textbf{x};t)$ can in its turn be  written approximately in
terms of $\overline{n}({\bf r};t)$ and
$\sigma(k;\textbf{r},t)$, thus introducing strong non-linear
effects. Thus, even before solving Eqs. (\ref{difeqdlp}) and
(\ref{relsigmadif2p}), they reveal a number of relevant features
of general and/or universal character.

The most illuminating of them is that, besides the equilibrium
stationary solutions $\overline{n}^{eq}(\textbf{r})$ and
$\sigma^{eq} (k;\textbf{r})$, defined by the equilibrium
conditions $\nabla \beta\mu[{\bf r};\overline{n}^{eq}]=0$ and
$\mathcal{E}(k;\overline{n}(\textbf{r},t))
\sigma(k;\textbf{r},t)=1$, Eqs. (\ref{difeqdlp}) and
(\ref{relsigmadif2p}) also predict the existence of another set of
stationary solutions that satisfy the dynamic arrest condition,
$\lim_{t\to \infty} b(\textbf{r},t)=0$. This far less-studied
second set of solutions describes, however, important
non-equilibrium stationary states of matter, corresponding to
common and ubiquitous non-equilibrium amorphous solids, such as
glasses and gels.

\subsection{Spatial uniformity, a simplifying approximation.}\label{subsection9.2}

To appreciate the essential physics of this fundamental and
universal prediction of Eqs. (\ref{difeqdlp}) and
(\ref{relsigmadif2p}), the best is to provide explicit examples.
To do this without a high mathematical cost, however, let us write
$\overline{n}(\textbf{r},t)$ as $\overline{n}(\textbf{r},t)=
\overline{n}(t) + \Delta \overline{n}(\textbf{r},t)$, and in a
first stage let us neglect the spatial heterogeneities represented
by the deviations $\Delta \overline{n}(\textbf{r},t)$. As a
result, rather than solving the time-evolution equation for
$\overline{n}({\bf r};t)$, we have that $\overline{n}(t)$ now
becomes a control parameter, so that we only have to solve the
time-evolution equation for the covariance
$\sigma(k,\textbf{r};t)$. We may consider, for example, the
specific case in which the system is {\it constrained} to remain
isochoric and spatially {\it homogeneous} ($\overline{n}({\bf
r};t)\approx \overline{n}\equiv N/V$) after an instantaneous
temperature quench at time $t=0$, from an arbitrary initial
temperature to a lower final temperature $T$. For this process,
the time-evolution equation for the Fourier transform (FT)
$\sigma(k;t)$ of the covariance $\sigma(\textbf{r},
\textbf{r}';t)=\sigma(\mid\textbf{r}-\textbf{r}'\mid;t)$ can
be written, for $t>0$ and in terms of the non-stationary static
structure factor $S(k;t)\equiv \sigma(k;t)/\overline{n}$, as

\begin{equation}
\frac{\partial S(k;t)}{\partial t} = -2k^2 D^0
b(t)\overline{n}\mathcal{E}_f(k) \left[S(k;t)
-1/\overline{n}\mathcal{E}_f(k)\right]. \label{relsigmadif2pp}
\end{equation}
in which $\mathcal{E}_f(k)=\mathcal{E}(k;\overline{n},T_f)$ is the Fourier
transform (FT) of the functional derivative
$\mathcal{E}[\mid\textbf{r}-\textbf{r}'\mid;n,T] \equiv \left[
{\delta \beta\mu [{\bf r};n]}/{\delta n({\bf r}')}\right]$ of the
chemical potential $\mu$, evaluated at $n({\bf r})=\overline{n}$
and $T=T_f$.

It is important to mention that the solution of this
equation yields in principle $S(k;t)$ as output, for given
$b(t)$ provided as input. This calls for an independent
relationship between these two unknowns, which may have the format
of an equation (or system of equations) that accepts $S(k;t)$ as
input and yields $b(t)$ as output. This is precisely the role of
the following set of equations. The first of them is an expression
for the time-evolving mobility $b(t)$,
\begin{equation}
b(t)= [1+\int_0^{\infty} d\tau\Delta{\zeta}^*(\tau; t)]^{-1},
\label{bdt}
\end{equation}
in terms of the $t$-evolving, $\tau$-dependent friction
coefficient $\Delta{\zeta}^*(\tau; t)$, which can be
approximated by \cite{nescgle1}
\begin{equation}
\begin{split}
  \Delta \zeta^* (\tau; t)= \frac{D_0}{24 \pi
^{3}\overline{n}}
 \int d {\bf k}\ k^2 \left[\frac{ S(k;
t)-1}{S(k; t)}\right]^2 \times \\
F(k,\tau; t)F_S(k,\tau; t).
\end{split}
\label{dzdtquench}
\end{equation}

In this equation $\tau$ is the \emph{correlation time} and $t$
is the \emph{waiting} (or evolution) time. $F(k,\tau; t)$ and
$F_S(k,\tau; t)$ are, respectively, the collective and self
non-equilibrium intermediate scattering functions (ISFs), whose
respective memory functions are approximated to yield the
following approximate expressions for the Laplace transforms (LT)
$\hat F(k,z; t)$ and $\hat F_S(k,z; t)$,
\begin{gather}\label{fluctquench}
\hat  F(k,z; t) = \frac{S(k; t)}{z+\frac{k^2D^0 S^{-1}(k;
t)}{1+\lambda (k;t)\ \Delta \hat  \zeta^*(z; t)}},
\end{gather}
and
\begin{gather}\label{fluctsquench}
\hat  F_S(k,z; t) = \frac{1}{z+\frac{k^2D^0 }{1+\lambda (k;t)\
\Delta \hat \zeta^*(z; t)}}.
\end{gather}
In these equations $\lambda (k)$ is a phenomenological
``interpolating function" \cite{nescgle1}, given by
\begin{equation}
\lambda (k;t)=1/[1+( k/k_{c}(t)) ^{2}], \label{lambdadk}
\end{equation}
with $k_c(t)$ being an empirically chosen cutoff wave vector. Eqs. (\ref{dzdtquench})-(\ref{lambdadk}) are the non-equilibrium
extension of the corresponding equations of the equilibrium SCGLE
theory, which is recovered in the long-$t$ stationary limit in
which $S(k;t\to\infty) \to S^{(eq)}(k) \equiv
1/\overline{n}\mathcal{E}_f(k)$. The derivation of these equations
in Ref. \cite{nescgle1} also extends to non-equilibrium
conditions the same approximations and assumptions employed in the
original derivation of the \emph{equilibrium} SCGLE theory \cite{todos2}.
The development of such an extension was quite natural within the framework of the
non-equilibrium generalization of Onsager's theory, but it is not in the
context of the Mori-Zwanzig  formalism \cite{boonyip},  which is deeply rooted in
the equilibrium condition.

Eqs. (\ref{relsigmadif2pp})-(\ref{lambdadk}) constitute the core of the simplest version of the NE-SCGLE  theory of irreversible processes in liquids. Their simultaneous self-consistent solution  provides the NE-SCGLE  description of the spontaneous evolution of the structure and dynamics of an \emph{instantaneously} and \emph{homogeneously} quenched \emph{monocomponent} liquid. A concrete application of this theory starts with the selection of the specific physical system whose non-equilibrium properties are to be modeled. In its present spatially-uniform version, which does not consider external fields, such specific system is represented by its pair interaction potential $u(r)$. One must then determine the fundamental  thermodynamic relation (FTR) $S=S[n;T]$, which describes the dependence of the entropy functional $S[n;T]$ on the density field $n(\mathbf{r})$. The FTR is in principle determined by the exact Boltzmann's principle $S[n;T]=k_B \ln W[n;T]$ or by any approximate method of equilibrium statistical mechanics \cite{mcquarrie,hansen}. This provides the thermodynamic input of the NE-SCGLE theory (i.e., Eqs. (\ref{relsigmadif2pp})-(\ref{lambdadk}) above). The solution of these equations for arbitrary initial condition $S_i(k)\equiv S(k;t_w=0)$ yields the predicted non-equilibrium evolution of the  $t_w$-dependent structure factor $S(k;t_w)$ and dynamic properties, such as $F(k,\tau; t_w)$ and $F_S(k,\tau; t_w)$.

\subsection{Brief review of concrete solutions of the NE-SCGLE equations.}\label{subsection9.3}

Although the NE-SCGLE theory was put forward a bit more than one decade ago, the canonical protocol just described has already been followed for a rich variety of systems, predicting an amazing diversity of features associated with the non-equilibrium relaxation of instantaneously quenched liquids. The most striking prediction of the solution of the NE-SCGLE equations (\ref{relsigmadif2pp})-(\ref{lambdadk}), is the revelation of the existence of the non-equilibrium stationary solutions that correspond to the formation of non-equilibrium solid materials as a result of the amorphous solidification of glass- and gel-forming liquids. Thus, for example, for  simple liquids with purely repulsive interactions, this theory provided a detailed description of the  non-stationary and non-equilibrium transformation of equilibrium hard- (and soft)-sphere liquids, into ``repulsive'' (high-temperature, high-density) hard-sphere glasses \cite{nescgle3,nescgle4}. These predictions have recently found a reasonable agreement with the corresponding non-equilibrium simulations  \cite{nescgle6},  naturally explaining some of the most essential signatures of the glass transition \cite{angellreview1,ngaireview1}. This success has rapidly branched in several relevant and surprising directions.

As a very brief review, let us describe some of these recent developments. For example, when adding attractions to the purely repulsive interactions just referred to, the same theory predicts new dynamically-arrested phases, identified with gels and porous glasses \cite{nescgle5}, and provides a kinetic perspective of the irreversible evolution of the structure of the system after being instantaneously quenched to the interior of its spinodal region \cite{nescgle7,nescgle8}. It also explains the otherwise seemingly complex interplay between spinodal decomposition, gelation, glass transition, and their combinations \cite{sastry, cocard, wyss, zaconedelgado, guoleheny, coniglio, chaudhuri2}.

This has recently led \cite{beni1} to a sound theoretical definition, and first-principles prediction, of the concept of non-equilibrium phases and time-dependent (i.e., aging) phase diagrams, a step forward in the understanding and \emph{prediction} of the  fascinating  non-equilibrium transient states observed in the formation of many soft and hard materials from the cooling of liquids and melts, quite commonly represented in rather unorthodox ``phase diagrams''  in which the (``waiting'') time after preparation becomes a key variable. Practical examples range from gel or glass formation by clays \cite{ruzicka1,ruzicka2} or proteins \cite{cardinaux,gibaud} in  solution, to the experimental phase transformations in the manufacture of hard materials, such as iron and steel \cite{callisterrethwisch,atlasttd} and porous glasses \cite{nakashima}.

On the other hand, extended to  multi-component systems \cite{nescgle4,nescgle6}  the NE-SCGLE theory opens the possibility of describing the aging of  ``double'' and ``single'' glasses in mixtures of neutral \cite{voigtmanndoubleglasses,rigo,lazaro} and charged \cite{prlrigoluis,portadajcppedro} particles; the initial steps in this direction are highly encouraging \cite{expsimbinarymixt2017}. Similarly, its extension to liquids formed by particles interacting by non-radially symmetric forces \cite{gory1,gory2}, accurately predicts the non-equilibrium coupled translational and rotational dynamic arrest observed in simulations \cite{gory3}.

Given these advances in the statistical physics of non-equilibrium liquids, a natural and immediate question refers to the relationship, and possible interplay, between the conventional statistical thermodynamic theory of equilibrium fluids  (represented, for example, by integral equation \cite{mcquarrie,hansen} or density functional theory \cite{evans}) and this emerging non-equilibrium theory.  One explicit example of common interest is, of course, the challenge of determining the fundamental  thermodynamic relation (FTR) $S=S[n;T]$
in any new application. Determining $S[n,T]$ is a classical and relevant  \emph{goal} of equilibrium statistical thermodynamics, whereas $S[n,T]$ is an essential \emph{input} in each specific application of the NE-SCGLE theory. Most likely, such common interests will stimulate  a creative communication between the practitioners of these two areas of statistical physics.



\section{Discussion and perspectives.}\label{section7}\label{section6}


In this work we have reviewed the use of simple mathematical models of stochastic dynamics as the backbone of relevant physical theories. We started in Section \ref{section2} with Langevin's theory of Brownian motion and  the Onsager-Machlup theory of equilibrium thermal fluctuations, whose mathematical backbone is the Ornstein-Uhlenbeck stochastic process. An early pioneering contribution of the Mexican statistical physics community was the proposal in 1987 of a non-Markov extension of the OU model, which allowed the inclusion of memory effects in the original Onsager-Machlup theory. The resulting non-Markov OM theory turned out to be mathematically  equivalent to the generalized Langevin equation, normally associated with Mori-Zwanzig projection operators employed in its derivation. 

As an application of the non-Markov Onsager-Machlup theory, it was possible to derive the exact results employed to construct an approximate statistical mechanical theory of the dynamic properties of equilibrium liquids, referred to as the SCGLE theory. This theory turned out to be essentially equivalent to mode coupling theory, correctly predicting the  existence of dynamically-arrested states of matter and the behavior of equilibrium viscous liquids near their glass transition. Exactly at the glass transition and beyond, however, both theories predict the strict vanishing of the particles' mobility and the divergence  of the viscosity and of the $\alpha$-relaxation time. Unfortunately, such singular behavior is never observed experimentally. 

Instead of such an abrupt and singular transition predicted by MCT and by the equilibrium SCGLE theory, what is experimentally observed is a smooth and continuous transition that changes with the time of observation in an extremely slow fashion. These so-called ``aging'' processes,  with their dependence on the protocol of fabrication of the material \cite{angellreview1,ngaireview1,anderson,sciortino}, constitute genuine fingerprints of glasses and the glass transition, whose molecular explanation is one essential aspect of a second long-standing fundamental challenge of  statistical physics, also successfully addressed by the Mexican statistical physics community. 

In order to formulate a first-principles statistical mechanical description of the glass transition, including the aging processes in glass- and gel-forming liquids during the formation of non-equilibrium amorphous solids, the need aroused to first extend the  (Markov and non-Markov) Onsager-Machlup theory to non-stationary, non-equilibrium, and non-linear conditions. This was done by first extending its mathematical backbone, i.e., the non-Markov Ornstein-Uhlenbeck stationary stochastic process, to non-stationary conditions. This resulted in the  definition of the piece-wise stationary model of globally non-stationary but locally stationary stochastic processes. This mathematical model was presented in Section \ref{section3}, and its use as the mathematical backbone of the general theory of thermal fluctuations in non-equilibrium systems, that we refer to as the \emph{non-equilibrium Onsager-Machlup theory,} was explained in Section \ref{section4}.

The material discussed in Section \ref{section2} was meant to be a pedagogical review of textbook material, including an illustrative application (the derivation of the equilibrium SCGLE theory). To a large extent, sections \ref{section3} and \ref{section4} might also seem to be just a review of the original proposal of a non-equilibrium extension of the OM theory, first presented in Refs. \cite{nescgle00,nescgle0}. This, however, is not the case. What makes a profound difference is the merging we have made in section \ref{section3}, of the concept of piecewise stationary process, already present in Refs. \cite{nescgle00,nescgle0}, and the concept of Lyapunov stability, as an explicit additional condition in the new definition of the PS stochastic process. Although  in practice this addition does not introduce noticeable formal changes, it provides a fundamental justification for the determination of the $t$-dependent local stationary covariance ${\bm{\sigma}}^{ss}(t)$ by means of the general result in Eq. (\ref{sigmaxvsdt}), ${\bm{\sigma}}^{ss}(t) \cdot  \mathcal{E}(t) = {\bm I}$, where  $\mathcal{E}(t)$ is defined as $\mathcal{E}(t)\equiv \mathcal{E}[\overline{\textbf a}(t)]$, with the stability matrix function $ \mathcal{E}[{\textbf a}]$ defined in Eq. (\ref{vquad}).

With the mathematical and physical infrastructure established in Sections \ref{section3} and \ref{section4}, we could finally describe in Section \ref{section5} the application of the abstract and general non-equilibrium Onsager-Machlup framework, to the formulation of a generic theory of the non-equilibrium relaxation of the structural and dynamical properties of a simple liquid subjected to the simplest manipulation or preparation protocol, namely, an instantaneous quench or compression. This resulted in Eqs. (\ref{relsigmadif2pp})-(\ref{lambdadk}) that constitutes the core of the simplest version of the NE-SCGLE  theory of irreversible processes in liquids. The simultaneous self-consistent solution of these equations provides the NE-SCGLE description of the spontaneous evolution of the structure and dynamics of our \emph{instantaneously} and \emph{homogeneously} quenched \emph{monocomponent} liquid.

During the last decade, the solution of the non-linear equations (\ref{relsigmadif2pp})-(\ref{lambdadk}) has  unveiled an amazing predicted scenario, involving two competing  kinetic pathways: the processes of thermodynamic equilibration, and the ultra-slow dynamic arrest processes leading to the formation of non-equilibrium amorphous solids. As briefly reviewed at the end of the last section, the stationary  solutions of these time-evolution equations are naturally grouped in two mutually-exclusive sets: the well-understood universal catalog  formed by all the states that maximize the entropy, and the newly-discovered, complementary universal catalog, of non-equilibrium amorphous states of matter. 

We must say, however, that in spite of how rich and enlightening this scenario predicted by the NE-SCGLE equations (\ref{relsigmadif2pp})-(\ref{lambdadk}) might seem, what we are witnessing is in reality only the emergence of a qualitatively new, non-equilibrium extension of the theory of liquids. The major fundamental contribution of the NE-SCGLE theory is its prediction of the existence of the second universal catalog of states of matter, constituted by the non-equilibrium dynamically-arrested states in which many materials can be found, from extremely common ones, like ordinary glasses and gels \cite{mauro,yoshi}, to subtle and complex non-equilibrium congested structures in soft and biological systems  \cite{komura}, and in deep-tech materials for space exploration, sustainable energy development, and biotechnology  \cite{liu}. 

Although the existence of this universal catalog of non-equilibrium states was first proposed by mode coupling theory \cite{goetze4}, its identification with the set of all the stable non-equilibrium stationary solutions of the NE-SCGLE equations (\ref{relsigmadif2pp})-(\ref{lambdadk}) introduces an essential kinetic perspective in the definition of the physical state of these materials. In its present, simplest version, the NE-SCGLE equations only have these two disjoint catalogs of states of matter, namely, the catalog of non-equilibrium stationary solutions and the good old catalog of equilibrium stationary states that maximize the entropy. 

In a real material, however,  different regions of the same macroscopic sample may be in different equilibrium or non-equilibrium states, a condition not yet contemplated in the present, simplest version, the NE-SCGLE theory. This is an example of the list of  areas of opportunity for improvement of the NE-SCGLE theory. This list, however, is actually enormous, and will surely be the subject of intense activity in the future. The sequence of successful applications briefly reviewed at the end of the last section, and the corresponding remarkable predictions, clearly point to a stimulating scientific journey to come.

\medskip
ACKNOWLEDGMENTS: We acknowledge Ana Gabriela Carretas-Talamante, Paulina Alejandra Ojeda-Mart\'inez, and Orlando Joaqu\'in-Jaime, for many valuable discussions. We are especially indebted to Diego Raid Peredo-Ortiz for his invaluable discussions on stability and dynamical systems, which inspired one of the fundamental aspects of the present work. This work was supported by the Consejo Nacional de Ciencia y Tecnolog\'ia (CONACYT, Mexico) through Postdoctoral Fellowships Grants No. I1200/224/2021 and I1200/320/2022; and trough grants 320983, CB A1-S-22362, and LANIMFE 314881.

\end{document}